\documentclass[twocolumn]{aastex631}

\usepackage{graphicx}	\usepackage{amsmath}	\usepackage{multirow}   \usepackage{booktabs}
\usepackage{rotating}

\usepackage{placeins}
\usepackage{float}
\newcommand{\oiii}{[\hbox{{\rm O}\kern 0.1em{\sc iii}}]\,5007}

\newcommand{\nii}{[\hbox{{\rm N}\kern 0.1em{\sc ii}}]\,6583}
\newcommand{\sii}{[\hbox{{\rm S}\kern 0.1em{\sc ii}}]\,6717,6731}
\newcommand{\siii}{[\hbox{{\rm S}\kern 0.1em{\sc iii}}]\,9069,9531}

\newcommand{\ciii}{[\hbox{{\rm C}\kern 0.1em{\sc iii}}]\,1907,1909}
\newcommand{\civ}{\hbox{{\rm C}\kern 0.1em{\sc iv}}\,1550}
\newcommand{\oiiiuv}{[\hbox{{\rm O}\kern 0.1em{\sc iii}}]\,1660,1666}
\newcommand{\oiiite}{[\hbox{{\rm O}\kern 0.1em{\sc iii}}]\,4363}
\newcommand{\heii}{\hbox{{\rm He}\kern 0.1em{\sc ii}}\,1640}
\newcommand{\hii}{\hbox{{\rm H}\kern 0.1em{\sc ii}}}

\newcommand{\oii}{[\hbox{{\rm O}\kern 0.1em{\sc ii}}]\,3727}
\newcommand{\hb}{\hbox{{\rm H}\kern 0.1em{\sc $\beta$}}}
\newcommand{\halpha}{\hbox{{\rm H}\kern 0.1em{\sc $\alpha$}}}

\newcommand{\ciiradio}{[\hbox{{\rm C}\kern 0.1em{\sc ii}}]$_{\rm 158 \mu m}$}

\newcommand{\SiII}{[\hbox{{\rm Si}\kern 0.1em{\sc ii}}]}

\newcommand{\OI}{[\hbox{{\rm O}\kern 0.1em{\sc i}}]}

\newcommand{\SiIIAbsA}{\hbox{{\rm Si}\kern 0.1em{\sc ii}$\,\lambda1190$}}
\newcommand{\SiIIAbsB}{\hbox{{\rm Si}\kern 0.1em{\sc ii}$\,\lambda1260$}}
\newcommand{\SiIIAbsC}{\hbox{{\rm Si}\kern 0.1em{\sc ii}$\,\lambda1526$}}
\newcommand{\CIIAbs}{\hbox{{\rm C}\kern 0.1em{\sc ii}$\,\lambda1334$}}

\newcommand{\SiIVAbs}{\hbox{{\rm Si}\kern 0.1em{\sc iv}$\,\lambda\lambda1393,1402$}}
\newcommand{\CIVAbs}{\hbox{{\rm C}\kern 0.1em{\sc iv}$\,\lambda\lambda1548,1550$}}

\newcommand{\M}{$\log_{10}(M_*/{\rm M}_{\odot})$}

\newcommand{\siion}{\ensuremath{\xi_{\rm ion}}}

\newcommand{\fesc}{$f\rm{_{esc}}$}
\newcommand{\Nion}{$N\rm{_{ion}}$}
\newcommand{\SF}{$\Delta\log_{10}(\mathrm{SFR}/M_\odot\,\mathrm{yr}^{-1})$}

\begin{document}

\title[Short title, max. 45 characters]{Exploring biases in derived stellar parameters and the ionizing photon production efficiency.}

\author[0000-0003-2035-3850]{Ravi Jaiswar}
\affiliation{International Centre for Radio Astronomy Research, Curtin University, Bentley WA, Australia}
\affiliation{ARC Centre of Excellence for All Sky Astrophysics in 3 Dimensions, Bentley WA, Australia}

\author{Anshu Gupta}
\affiliation{International Centre for Radio Astronomy Research, Curtin University, Bentley WA, Australia}
\affiliation{ARC Centre of Excellence for All Sky Astrophysics in 3 Dimensions, Bentley WA, Australia}

\author{Elisabete da Cunha}
\affiliation{International Centre for Radio Astronomy Research, University of Western Australia, Crawley WA, Australia}
\affiliation{ARC Centre of Excellence for All Sky Astrophysics in 3 Dimensions, Bentley WA, Australia}

\author{Cathryn M. Trott}
\affiliation{International Centre for Radio Astronomy Research, Curtin University, Bentley WA, Australia}
\affiliation{ARC Centre of Excellence for All Sky Astrophysics in 3 Dimensions, Bentley WA, Australia}

\author{Andrew Battisti}
\affiliation{International Centre for Radio Astronomy Research, Curtin University, Bentley WA, Australia}
\affiliation{ARC Centre of Excellence for All Sky Astrophysics in 3 Dimensions, Bentley WA, Australia}
\affiliation{Research School of Astronomy and Astrophysics, Australian National University, Cotter Road, Weston Creek, ACT 2611, Australia}

\author{A.J. Hedge}
\affiliation{International Centre for Radio Astronomy Research, Curtin University, Bentley WA, Australia}

\author{Robin Cook}
\affiliation{International Centre for Radio Astronomy Research, Curtin University, Bentley WA, Australia}

\author{Sabine Bellstedt}
\affiliation{International Centre for Radio Astronomy Research, Curtin University, Bentley WA, Australia}
\affiliation{ARC Centre of Excellence for All Sky Astrophysics in 3 Dimensions, Bentley WA, Australia}

\author{Jordan D'Silva}
\affiliation{International Centre for Radio Astronomy Research, Curtin University, Bentley WA, Australia}

\author{Luke Davies}
\affiliation{International Centre for Radio Astronomy Research, Curtin University, Bentley WA, Australia}
\affiliation{ARC Centre of Excellence for All Sky Astrophysics in 3 Dimensions, Bentley WA, Australia}

\author{Juno Li}
\affiliation{International Centre for Radio Astronomy Research, Curtin University, Bentley WA, Australia}
\affiliation{ARC Centre of Excellence for All Sky Astrophysics in 3 Dimensions, Bentley WA, Australia}

\begin{abstract}
Constraining the timescale and manner in which the Epoch of Reionization (EoR) occurred is a major JWST science goal. However, any constraints on the stellar or ionizing parameters (\siion) of galaxies in the EoR must contend with biases introduced by both the data and the models used. We explore three techniques that use spectroscopic and photometric data as well as three different spectral energy distribution (SED) fitting codes, each comprised of multiple star formation history, stellar population synthesis, dust, and photoionization prescriptions to determine their relative influence on stellar parameters and \siion . We use z$\sim$3 EoR analog galaxies due to their reliable photometric coverage (improved physical constraints) in comparison to direct EoR sources and potential for direct Lyman Continuum escape research. For this population the median stellar mass can vary by over 0.6 dex and the SFR by more than 0.9 dex. Further, the \siion\, can vary by over 1.1 dex for individual sources when comparing spectroscopic and photometric derivations, or by more than 0.5 dex when fitting SEDs with different models. As such, the choice of methodology can have significant consequences for the derived \siion\, and the subsequent sources of reionization. We find that the presence of a redshift evolution for \siion\, is dependent on the method adopted for its derivation, where a consistent method yields an evolutionary trend with redshift  in extreme emitters while an indiscriminant selection of studies does not. The model, method and data dependence of the \siion\, parameter is undeniable even for a homogeneous population.
\end{abstract}

\section{Introduction} \label{sec:intro}
The Epoch of Reionization (EoR) refers to the period following the first formation of stars and galaxies, where the neutral hydrogen gas underwent a dramatic phase shift in the intergalactic medium (IGM) \citep{Robertson2022}. Massive O and B stars from primordial galaxies produced strongly ionizing Lyman Continuum (LyC) radiation; some of which managed to escape the galaxy and traverse the IGM, ionizing its neutral hydrogen gas. This process transformed the predominantly neutral universe to be largely ionized today.  

The specific timescale over which the early universe was ionized is still a matter of debate. Models favoring the contribution of massive bright galaxies create a sharper ionization rate curve, while those considering smaller sources take longer but start earlier \citep{Cook2024}. Active Galactic Nuclei have fallen out of favor as the primary source of ionizing radiation due to their infrequency at z$>$3 \citep{Kulkarni2018}. The current best estimates place the end of reionization at z$\sim$5.3 \citep{Bosman2022}. Determining the finer details requires an understanding of the production rate of ionizing radiation (\Nion) relative to the rate of ion recombination. The quest for \Nion\ requires precise measurements of the ionizing photon production efficiency (\siion) as well as the escape fraction of Lyman continuum light (\fesc) and the number density of these emitters ($\mu_\mathrm{UV}$), which together describe the amount of ionizing radiation entering the IGM \citep{Robertson2013}.  
\begin{figure}
    \centering
    \includegraphics[width=0.95\linewidth]{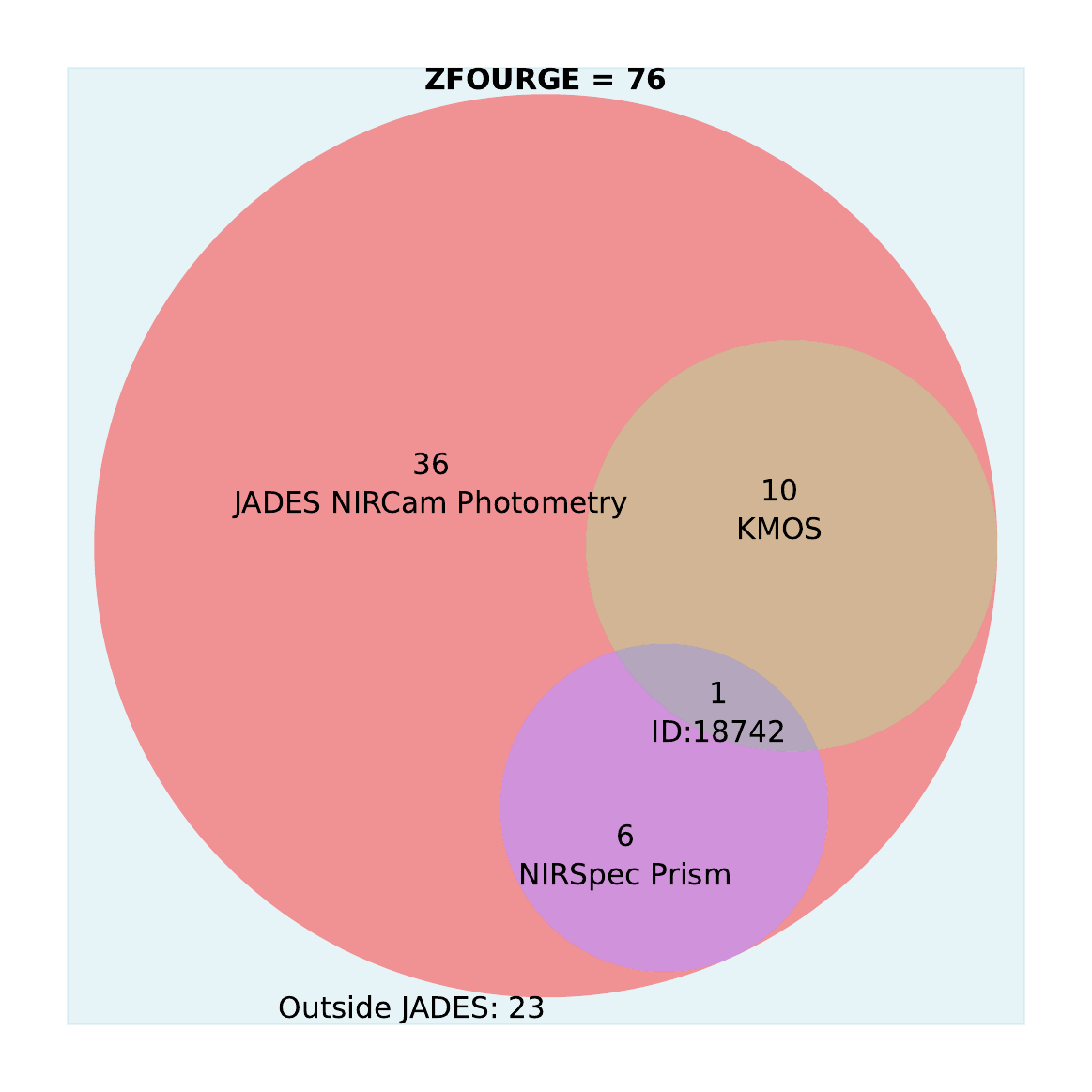}
    \caption{Sample Venn diagram indicating original EELG sample (76 sources, \citep{Forrest2018, Jaiswar2024}) JADES \citep{Eisenstein2023,Eisenstein2023a,Rieke2023,Hainline2023}) NIRCam photometry footprint overlap (53 sources- JESCO sample) and two spectroscopic subsamples \citep{Tran2020,Gupta2022},\citep{Bunker2023}.}
    \label{fig:Data}
\end{figure}

Determining \siion\, has been the goal of multiple studies relating galactic properties to the evolution of the IGM in the EoR \citep[e.g.][]{Begley2024,Pahl2024,Vanzella2024,Zhu2024}. In the era of JWST, it is now practical to obtain well-sampled data for statistical analysis of high redshift star-forming galaxies \citep{Bagley2022,Bunker2023,Oesch2023,Williams2023,Tran2023,Llern-ref,Begley-ref,Pahl-ref}, from which to probe \siion\, through both wide-coverage SED fitting and spectral analysis, setting the stage for stricter constraints.

However, the current landscape of research into \siion\, is overwhelmingly inconsistent in its methods and findings. Sources at increasingly impressive redshifts reveal variability in \siion\, values $\Delta$log$_{10}$(\siion)$\sim0.4$ dex between studies at z=10-12 \citep{Hsiao2024,Calabro2024} and $\Delta$log$_{10}$(\siion)$\sim0.7$ dex at z=6-7 \citep{Begley2024,Vanzella2024}, resulting in vastly different final reionization timescales. Larger population analyses echo this, with \cite{Simmonds2024} sources covering a wide dynamical range by z$\sim$8, and studies even finding $\Delta$log$_{10}$(\siion)$\sim0.6$ dex for the exact same sources using different models at z$\sim$8 \citep{Tang2023,Jaiswar2024}. 

Furthermore, the existence of a redshift evolution for \siion\, is still debated. \cite{Castellano2023,Pahl2024,Zhu2024} find no evolution while \cite{Rinaldi2023,Chen2024} find it to be significant. Young massive stars are thought to be more prevalent in the early universe due to the low metallicity nebulae \citep{Bromm2011,Maiolino2024}, or a top-heavy IMF \citep{Cameron2023} which could lead to a higher  production of ionizing photons at a given total stellar mass, but the argument of a mass bias is also a consideration \citep{Simmonds2024}. Additionally, at high redshift, newly formed and merging galaxies may contain morphological holes in the ISM \citep{Ji2020,Kerutt2024,Gupta2024}, which could alter the primary mode of LyC escape when compared to sources at low redshift \citep{Izotov2018}. 

Lower redshift analogs are often selected to study \siion\, \citep{Tang2019}, due to the improved physical constraints available by the additional photometric coverage of the rest optical and NIR portions of the SED. At z$<$4, these sources have potential for direct f$_{esc}$ studies as the density of neutral gas is much lower. \cite{Jaiswar2024} showed that a selection of Extreme \oiii\ emitters at z$\sim$3 matches well to direct EoR galaxies in terms of their physical properties. 

\begin{figure*}
    \centering
    \includegraphics[width=0.4\linewidth]{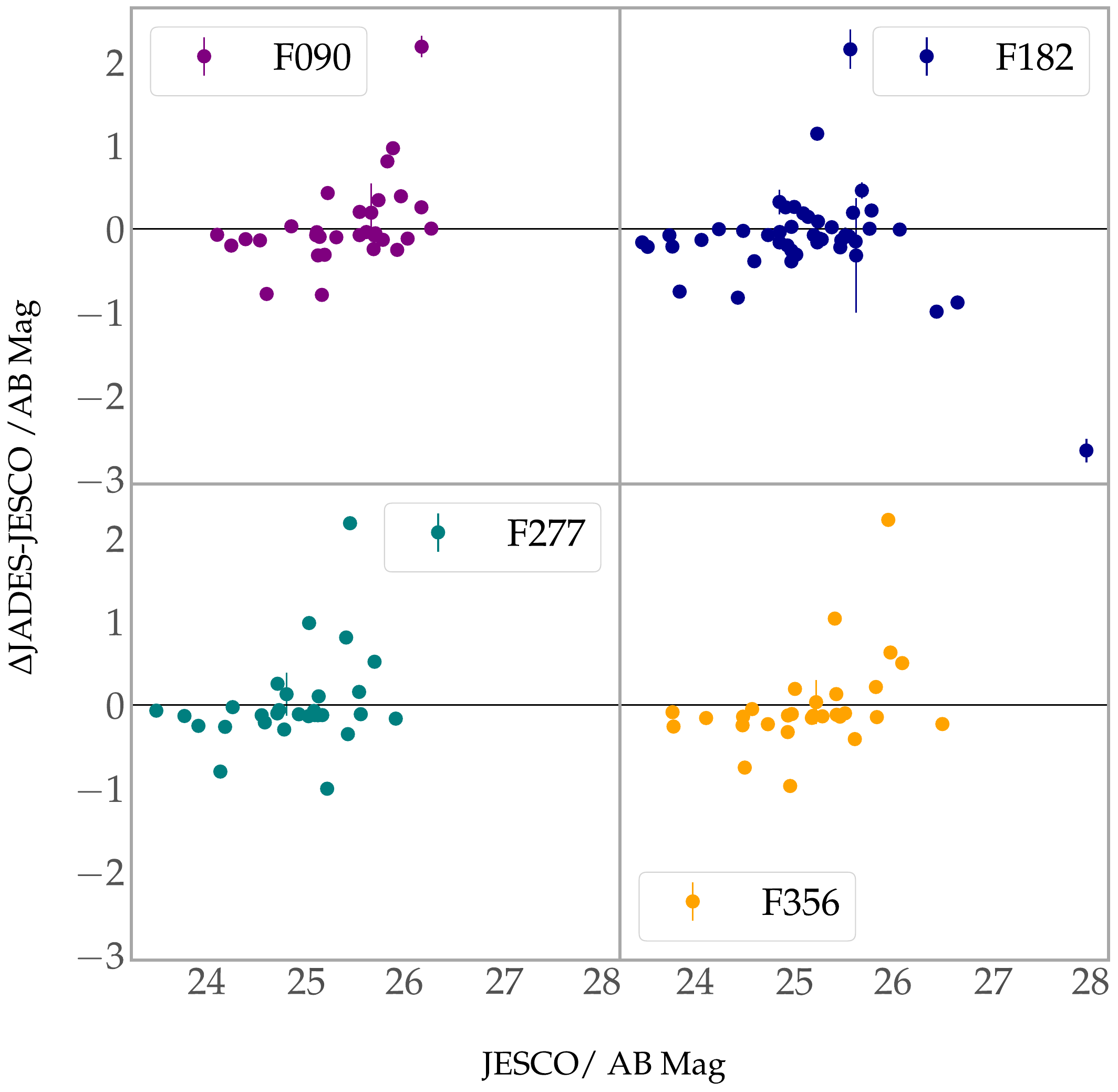}
    \includegraphics[width=0.59\linewidth]{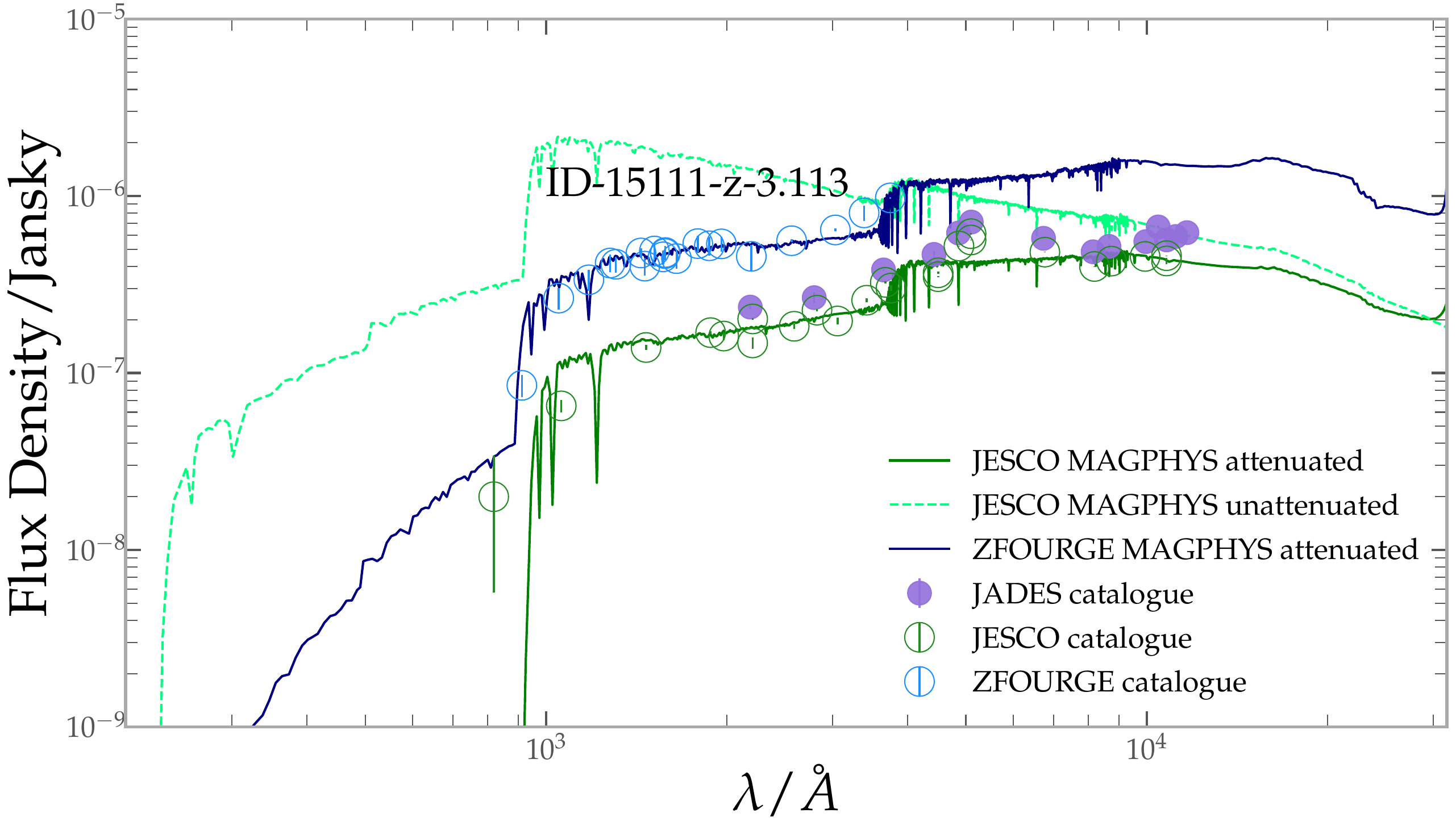}
    \caption{(Left) Flux density discrepancy in four JWST filters between the JADES survey and the JESCO catalogue which is described in section \ref{sec:data}. Zero discrepancy is marked by the black horizontal line. (Right) \texttt{MAGPHYS} SED of an example galaxy at z=3.545 comparing the fit using the ZFOURGE photometry (blue empty circles) to the JESCO photometry (green empty circles). The JADES photometry (purple full circles) for this galaxy is also presented for comparison. The ZFOURGE photometry is offset by an aperture correction factor due to the unique method used in its reduction \citep{Straatman2016}.}
    \label{fig:zfourge}
\end{figure*}

While differences in \siion, based on individual sources, could be attributed to variance in the samples, this viewpoint does not account for implicit biases in the techniques used by different studies. In addition, confirming the existence of a redshift evolution or any other correlation requires that a consistent method be used at each epoch while considering the most appropriate model assumptions for a specific population. Biases may be introduced through the comparison of different models and assigned to the galaxy itself erroneously. Addressing the consistency of different dust and star formation models in deriving \siion\, and stellar population parameters is, therefore, a challenging endeavor.

Spectral energy distribution models are a framework through which we interpret the emission from numerous astronomical sources and act to translate photometry into physical properties \citep{DaCunha2008,DaCunha2015,Chevallard2016,Carnall2017,Robotham2020,Johnson2021}. For galaxies, these are capable of using the photometry at different wavelengths to infer the mass, composition, and formation history of different phases of matter (stars, dust, gas). A single SED fitting code is itself comprised of multiple models governing the different phases of matter or sections of the SED simultaneously \citep{Conroy2013}. Fundamentally, deriving the stellar properties requires star formation history \citep{Carnall2019,Robotham2020}, dust attenuation \citep{Calzetti1994,Zubko1999,Charlot2000,Chevallard2013,Salim2018,Battisti2022}, dust emission \citep{Dale2014} and stellar population synthesis \citep{Bruzual2003,Conroy2008,Gutkin2016,Eldridge2017} models, though additional considerations such as the IGM absorption \citep{Madau1995,Inoue2014}, Active galactic Nuclei  \citep{Siebenmorgen2015} or photoionization models \citep{Dopita1996,Ferland2017} can improve the extracted information depending on the aims of the work \citep{Pacifici2023}. Therefore, there are practically hundreds of viable model combinations that make sense on the surface for any study.    
 
Furthermore, the physical properties informed by SEDs depend on the wavelength coverage and depth. For instance, the Star Formation Rate (SFR) may be probed in the FIR to radio range, particularly at low redshift \citep{Kennicutt2012,Davies2013,Davies2015}  over longer timescales, or using rest-frame optical emission lines at high redshift for short timescales \citep{Calzetti2012} due to the sources and timescales of their production (dust re-emission vs synchrotron emission \cite{Condon1992, Murphy2011}). By combining SED fitting with photoionization models, SFR timescales below 20 Myr can be constrained, as young stars (ages $<$20 Myr) dominate the ionizing flux that produces nebular \halpha\ and \oii\ emission \citep{Davies2016}.

A compelling application of SED modeling in determining \siion\, is in its ability to correct for dust reddening in galactic emissions. Interstellar dust absorbs high-energy stellar radiation, attenuating this portion of the SED and re-emitting it in the infrared \citep{Kennicutt2012}. By modeling the attenuation parameter we can recover something resembling the intrinsic stellar population spectrum \citep{Smith2018}, gaining insights into the internal mechanics of the interstellar medium (ISM) and accurately estimating \siion\ .

Stellar population synthesis (SPS) models combine stellar isochrones, IMFs, chemical abundances, and stellar physics to describe the evolving populations that produce galaxy SEDs. As these models incorporate more complex or specialized physics, their consistency across studies requires careful consideration \citep{Bellstedt2024}. Modern SED fitting tools like \texttt{ProSpect} \citep{Robotham2020} allow flexibility in internal assumptions, such as the IMF, increasing potential discrepancies between works. While BC03 \citep{Bruzual2003} remains widely used, newer models like BPASS \citep{Eldridge2017} account for binary interactions, which can significantly affect the evolution of massive stars. Mass transfer and stellar rotation can reduce core pressures and extend stellar lifetimes, influencing key parameters such as the ionizing photon output \citep{Yates2023}.
 
In this paper we explore the effects of three SED models and their subcomponents on the derived stellar parameters, with a focus on the \siion\, for a sample of EoR analogs at z$\sim$3. Using a proven analog sample allows for the depth and coverage afforded by lower redshifts while remaining relevant to the implications for \siion\, studies in the EoR. 
Section \ref{sec:data} provides a breakdown of the sample chosen. Section \ref{sec:methods} explains the methodology in exploring the different SED codes and the different methods of deriving \siion\, which introduces a further disparity in the field. The results and their discussion can be found in Section \ref{sec:results} with a separate exploration of the impacts on the redshift evolution in Section \ref{sec:Redshift Evolution}. We leave some closing comments in Section \ref{sec:conclusions}. All magnitudes are in the AB magnitude system \citep{Oke1983}.

\section{Data} \label{sec:data}
\subsection{The Sample}
The sample studied in this work was originally identified in the Chandra Deep Field South (CDFS) field by \cite{Forrest2018} as having extreme \oiii\ ($>$400 \AA\ EW) with spectroscopic follow up in \cite{Tran2020} and \cite{Gupta2022}. \cite{Jaiswar2024} established these objects as an analogous sample to EoR galaxies in terms of their physical properties (See Figure \ref{fig:Data}). There, it was found that these galaxies exhibit extremely {blue median UV slopes $=-1.83^{+0.21}_{-0.15}$ and, when modeled with \texttt{MAGPHYS} \citep{DaCunha2008,DaCunha2015}, low median $\rm{A_V}$ $=0.25^{+0.12}_{-0.14}$ and median} $\rm{M_*}$$= 9.75^{0.20}_{-0.28}$. 

The original sample of 76 Extreme Emission line galaxies (EELGs) (refer to Figure \ref{fig:Data} for breakdown) was derived from the FourStar Galaxy Evolution Survey (ZFOURGE, \cite{Straatman2016}) which uses medium bandwidth filters to derive robust photometric redshifts ($\sim$2\%) accuracy redshifts. From these, the subsample which fell into the footprint of the JWST Advanced Deep Survey (JADES, \citealt{Bunker2023,Eisenstein2023,Eisenstein2023a,Rieke2023,Hainline2023}) NIRCam images was selected for this study (53 galaxies, $\rm{2.5<z<4}$) using the original ZFOURGE redshifts.{We found that the ZFOURGE redshifts matched the JADES redshifts where they were both available to within 8\% for all sources with a median of $<$2\%.} With the current availability of JWST data it is possible to fully sample the optical-NIR SED of galaxies at $z\sim3$ up to 1.2$\mu m$ with 10$\sigma$ depth of 29.89 magnitude in F444W \citep{Eisenstein2023}.   

This analysis targets $z\sim 3$ analogs due to the rest-optical HST+JWST NIRCam photometric coverage from $\sim0.09 - 1.2\mu m$, providing excellent constraints to both the young and old stellar populations (see Figure \ref{fig:zfourge}). The choice of analogs at this redshift ensures availability of the necessary rest UV photometry to constrain the O and B stellar population dominating ionizing photon production while ensuring some NIR coverage for dust attenuation and mass constraints. This choice also allows follow-up studies to pursue direct f$_{esc}$ measurements near the highest redshifts with theoretically observable LyC. 

\subsection{Photometry}
We combine the 14 deep NIRCam images from JADES which consist of medium-band imaging from the JWST Extragalactic Medium band Survey \citep[JEMS;][]{Williams2023}
with the First Reionization Epoch Spectroscopically Complete Observations (FRESCO) \citep[FRESCO;][]{Oesch2023}. We also include mosaics created from the archival HST imaging by the FRESCO team \citep{Oesch2023} to create a photometric catalog with 35 filters which we refer to henceforth as the JESCO (JADES+FRESCO) catalog (29 unique filters with the deeper imaging taking priority). 

This catalog is produced using the ProFound software to extract the photometry \citep{Robotham2018}, where the segmentation image was created on a stack of the F182M and F444W images from the JADES survey wherever possible, otherwise substituting the equivalent FRESCO image or F160W when neither was possible (affecting only 3 galaxies). Stacking was performed to determine segments on both the stars and gas components of the population. Sources within 0.5 arcseconds ($< 10$ kpc at $z\sim3$) of each other which appear in the same filters in the RGB image (B:F182M, G:F210M, R:F444W) indicating similar redshifts were considered as a single source. Our derived fluxes were evaluated against the JADES DR1/2 catalogs, where a flux density match was found in 52/53 cases, with the remaining galaxy differing due to the newly defined segment considering nearby sources as merged. An example of this flux density match can be seen in Figure \ref{fig:zfourge}. We provide the derived photometry in Table \ref{tab:JESCO1} in a machine-readable format. This photometry is used throughout this study for every photometric experiment.

We also note that the new photometry derived in this work is around half the flux of the published ZFOURGE photometry. ZFOURGE survey implements an additional aperture correction based on the total flux lost out $1''.2$ aperture in K-band (see Figure \ref{fig:zfourge}). We suspect this aperture correction overestimates the total flux for EELGs with a typical size of $0''.5$, could be responsible for overestimated stellar mass and SFR in \cite{Jaiswar2024}, and can be seen in Figure \ref{fig:zfourge} (right). A representative SED comparing the JESCO, JADES and ZFOURGE photometry indicates the agreement between the newer catalogs.   

\subsection{Spectroscopy}\label{subsec:spectroscopy}
We consider the spectroscopic derivation of \siion\, (described in section \ref{subsec:siionmethod}) for the subsample of our sources with either observations from the  K-band Multi Object Spectrograph (KMOS, \citealt{Gupta2022}) IFU (11 sources) or JADES prism DR1 \citep{Bunker2023} galaxies (7 sources) matched to the JESCO catalog. {We find that the ZFOURGE and KMOS redshifts match within 4\% for all sources and within 2\% for all JADES/ZFOURGE matched sources.} We note also the coincidental overlap of one galaxy observed with both instruments with two different H$\beta$ fluxes resulting in a large flux discrepancy (see section \ref{subsec:spectrophotometry}. This is somewhat mitigated by considering the use of both IFU and prism spectra (see \cite{Bunker2023}, as well as the 5-20\% spectral IR flux calibration error noted in the KMOS user manual.

\section{Methods}\label{sec:methods}

\subsection{MAGPHYS}\label{subsec:magphys}
Multi-wavelength Analysis of Galaxy Physical Properties (\texttt{MAGPHYS}) is an SED fitting package that derives basic physical properties from a provided photometry \citep{DaCunha2008,DaCunha2015}. It does this by assembling a library of dust and stellar models at a predetermined redshift (i.e., we do not use the photo\_z fit version from \citealt{Battisti2019}), identifies close-matching models by  $\chi^2$ minimization, and derives a marginalized likelihood distribution of each physical parameter. We use the high\_z version \citep{DaCunha2015} that includes the BC03 stellar population models \citep{Bruzual2003}, \cite{Charlot2000} dust models, \cite{1983QJRAS..24..267H} gray body dust emission, \cite{Chabrier2003} IMF, \cite{Madau1995} IGM attenuation model, and an delayed exponentially declining star formation history model with random superimposed bursts throughout the last 2Gyr. Models are pre-generated, therefore SEDs generated with 0-multiple bursts may be selected depending on the $\chi^2$ fit. 

\begin{sidewaystable*}

\centering
\resizebox{\textwidth}{!}{\begin{tabular}{l l cccc | l cccc | l cccc}
    \toprule
    & \multicolumn{5}{c}{ProSpect} & \multicolumn{5}{c}{BEAGLE} & \multicolumn{5}{c}{MAGPHYS} \\
    \cmidrule(lr){2-6} \cmidrule(lr){7-11} \cmidrule(lr){12-16}
    &  & SFR$_{10}$ & SFR$_{100}$ & M$_*$  & \siion\ &  & SFR$_{10}$ & SFR$_{100}$ & M$_*$  & \siion\ &  & SFR$_{10}$ & SFR$_{100}$ & M$_*$  & \siion\ \\
    &  & M$_\odot$ yr$^{-1}$ & M$_\odot$ yr$^{-1}$ & M$_\odot$ & erg Hz$^{-1}$ &  & M$_\odot$ yr$^{-1}$ & M$_\odot$ yr$^{-1}$ & M$_\odot$ & erg Hz$^{-1}$ &  & M$_\odot$ yr$^{-1}$ & M$_\odot$ yr$^{-1}$ & M$_\odot$ & erg Hz$^{-1}$ \\
    \midrule
    \multirow{6}{*}{{BC03}} 
    & exp    &$0.98^{+0.53}_{-0.38}$ & $0.85^{+0.38}_{-0.32}$ & $9.05^{+0.20}_{-0.24}$ & $25.12^{+0.05}_{-0.02}$ & dtau  & $0.76^{+0.47}_{-0.09}$ & $0.72^{+0.47}_{-0.12}$ & $9.54^{+0.18}_{-0.15}$ & $25.33^{+0.00}_{-0.07}$ & dtau & $0.99^{+0.56}_{-0.63}$ & $0.90^{+0.47}_{-0.45}$ & $9.07^{+0.20}_{-0.19}$ & $25.17^{+0.04}_{-0.05}$ \\
    & exp+b   &$1.27^{+1.12}_{-0.84}$ & $0.71^{+0.51}_{-0.17}$ & $9.11^{+0.23}_{-0.24}$ & $25.21^{+0.02}_{-0.06}$ & dtau+b & $1.31^{+1.19}_{-0.65}$ & $0.60^{+0.42}_{-0.21}$ & $9.41^{+0.17}_{-0.42}$ & $25.54^{+0.02}_{-0.09}$ & dtau+b  & $0.96^{+0.90}_{-0.53}$ & $0.86^{+0.77}_{-0.34}$ & $9.09^{+0.18}_{-0.24}$ & $25.18^{+0.04}_{-0.05}$ \\
    & exp+p  &$1.05^{+0.64}_{-0.52}$ & $0.95^{+0.43}_{-0.47}$ & $9.17^{+0.16}_{-0.17}$ &$25.15^{0.04}_{0.03}$  & cal+b   &$1.59^{+1.47}_{-1.54}$ & $0.73^{+0.51}_{-0.12}$ & $9.12^{+0.52}_{-0.39}$ & $25.56^{+0.00}_{-0.15}$ &  &  &  &  &  \\
    & exp+b+p &$1.34^{+1.13}_{-0.96}$ & $0.75^{+0.63}_{-0.38}$ & $9.23^{+0.14}_{-0.23}$ &$25.35^{+0.05}_{-0.12}$  & uni+b   &$1.27^{+1.05}_{-0.52}$ & $0.57^{+0.31}_{-0.16}$ & $9.33^{+0.24}_{-0.42}$ & $25.54^{+0.02}_{-0.09}$ &  &  &  &  &  \\
    & tsnorm &$1.11^{+1.21}_{-0.82}$ & $1.02^{+0.89}_{-0.67}$ & $9.37^{+0.17}_{-0.21}$ & $25.08^{+0.15}_{-0.08}$ & smc+b   &$1.37^{+1.17}_{-0.87}$ & $0.60^{+0.37}_{-0.18}$ & $9.22^{+0.31}_{-0.38}$ & $25.54^{+0.02}_{-0.12}$ &  &  &  &  &  \\
    & const  &$0.76^{+0.56}_{-0.36}$ & $0.76^{+0.56}_{-0.36}$ & $9.16^{+0.21}_{-0.21}$ & $25.03^{+0.01}_{-0.00}$ &  &  &  &  &  &  &  &  &  &  \\
    \midrule
    \multirow{2}{*}{{BPASS}} 
    & exp    &$1.05^{+0.92}_{-0.62}$ & $0.85^{+0.55}_{-0.25}$ & $9.00^{+0.13}_{-0.21}$ & $25.18^{+0.02}_{-0.02}$ &  &  &  &  &  &  &  &  &  &  \\
    & exp+b   &$1.32^{+1.33}_{-0.76}$ & $0.81^{+0.51}_{-0.43}$ & $9.03^{+0.13}_{-0.23}$ & $25.21^{+0.02}_{-0.04}$ &  &  &  &  &  &  &  &  &  &  \\
    \midrule
    \multirow{5}{*}{{BC16-NP}} 
    &  &  &  &  &  & dtau   &$0.73^{+0.44}_{-0.18}$ & $0.74^{+0.40}_{-0.24}$ & $9.61^{+0.16}_{-0.17}$ & $25.29^{+0.00}_{-0.02}$ &  &  &  &  &  \\
    &  &  &  &  &  & dtau+b  &$1.30^{+1.16}_{-0.87}$ & $0.56^{+0.53}_{-0.15}$ & $9.22^{+0.35}_{-0.52}$ & $25.58^{+0.07}_{-0.07}$ &  &  &  &  &  \\
    &  &  &  &  &  & cal+b    &$1.10^{+1.30}_{-1.06}$ & $0.64^{+0.59}_{-0.17}$ & $9.30^{+0.40}_{-0.65}$ & $25.53^{+0.05}_{-0.80}$ &  &  &  &  &  \\
    &  &  &  &  &  & uni+b    &$1.31^{+0.97}_{-0.84}$ & $0.48^{+0.57}_{-0.03}$ & $8.90^{+0.64}_{-0.38}$ & $25.58^{+0.06}_{-0.09}$ &  &  &  &  &  \\
    &  &  &  &  &  & smc+b    &$1.28^{+1.05}_{-1.04}$ & $0.54^{+0.48}_{-0.18}$ & $8.89^{+0.73}_{-0.52}$ & $25.58^{+0.02}_{-0.11}$ &  &  &  &  &  \\
    \midrule
    \multirow{5}{*}{{BC16-P}} 
    &  &  &  &  &  & dtau &$0.68^{+0.36}_{-0.18}$ & $0.68^{+0.37}_{-0.18}$ & $9.44^{+0.22}_{-0.13}$ & $25.24^{+0.01}_{-0.01}$ &  &  &  &  &  \\
    &  &  &  &  &  & dtau+b &$1.13^{+1.15}_{-0.75}$ & $0.42^{\lesssim-0.07}_{-0.004}$ & $8.97^{+0.32}_{-0.50}$ & $25.55^{+0.07}_{-0.08}$ &  &  &  &  &  \\
    &  &  &  &  &  & cal+b    &$1.29^{+1.12}_{-0.81}$ & $0.59^{+0.42}_{-0.12}$ & $9.21^{+0.17}_{-0.50}$ & $25.53^{+0.00}_{-0.04}$ &  &  &  &  &  \\
    &  &  &  &  &  & uni+b    &$1.43^{+1.27}_{-1.19}$ & $0.65^{+0.34}_{-0.25}$ & $9.48^{+0.31}_{-0.64}$ & $25.53^{+0.05}_{-0.06}$ &  &  &  &  &  \\
    &  &  &  &  &  & smc+b    &$1.40^{+1.10}_{-0.97}$ & $0.66^{+0.22}_{-0.18}$ & $9.49^{+0.30}_{-0.47}$ & $25.53^{+0.06}_{-0.02}$ &  &  &  &  &  \\
    \bottomrule
    
\end{tabular}}

\caption{Comparison of parameter IQRs for our 53 EELG sample across different models. The columns include star formation rates (log$_{10}$(SFR$_{10}$) and log$_{10}$(SFR$_{100}$)), stellar mass (\M), and the ionizing photon production efficiency parameter (log$_{10}$(\siion/$\rm{erg^{-1}~Hz}$)). We refer to the star formation history models as exp (exponentially declining), dtau (delayed exponentially declining), tsnorm (truncated skewed Normal) and const (constant). The dust models used are referred to as cal (Calzetti), uni (Universal) and smc (SMC) where the Charlot and Fall model is used by default in the dtau models. The +b refers to the inclusion of a burst (where in BEAGLE it refers to the inclusion of the CSFH free parameter), and the +p refers to the inclusion of the MAPPINGS photoionization model in ProSpect. The SPS model and whether it includes photoionization (P) or does not (NP) is provided for each sample.}
\label{tab:big_table}

\end{sidewaystable*}

For the sake of comparison, we have also incorporated custom SFHs computed without the inclusion of bursts. We also modified the stellar libraries to more finely sample the high redshift timescales, and lowered the dust attenuation prior distribution from its default value centered at $\mathrm{A_V\sim1}$ by 50\%. {The} \texttt{MAGPHYS} {`high z' code extends to higher $\mathrm{A_V}$ as it was built for IR selected sub-mm galaxies }\citep{DaCunha2015}.{ We adopted this lower prior as our galaxies were not IR selected, though the range still reaches the same limits.} For the calculation of the \siion\, parameter, we remove the effects of the IGM absorption model on the output SED. We use \texttt{MAGPHYS} as the standard against which the other more flexible/complex models are compared to in sections \ref{subsec:SFH} and \ref{subsec:Bursts}.
\begin{figure*}
    \centering
    \includegraphics[width=1.0\linewidth]{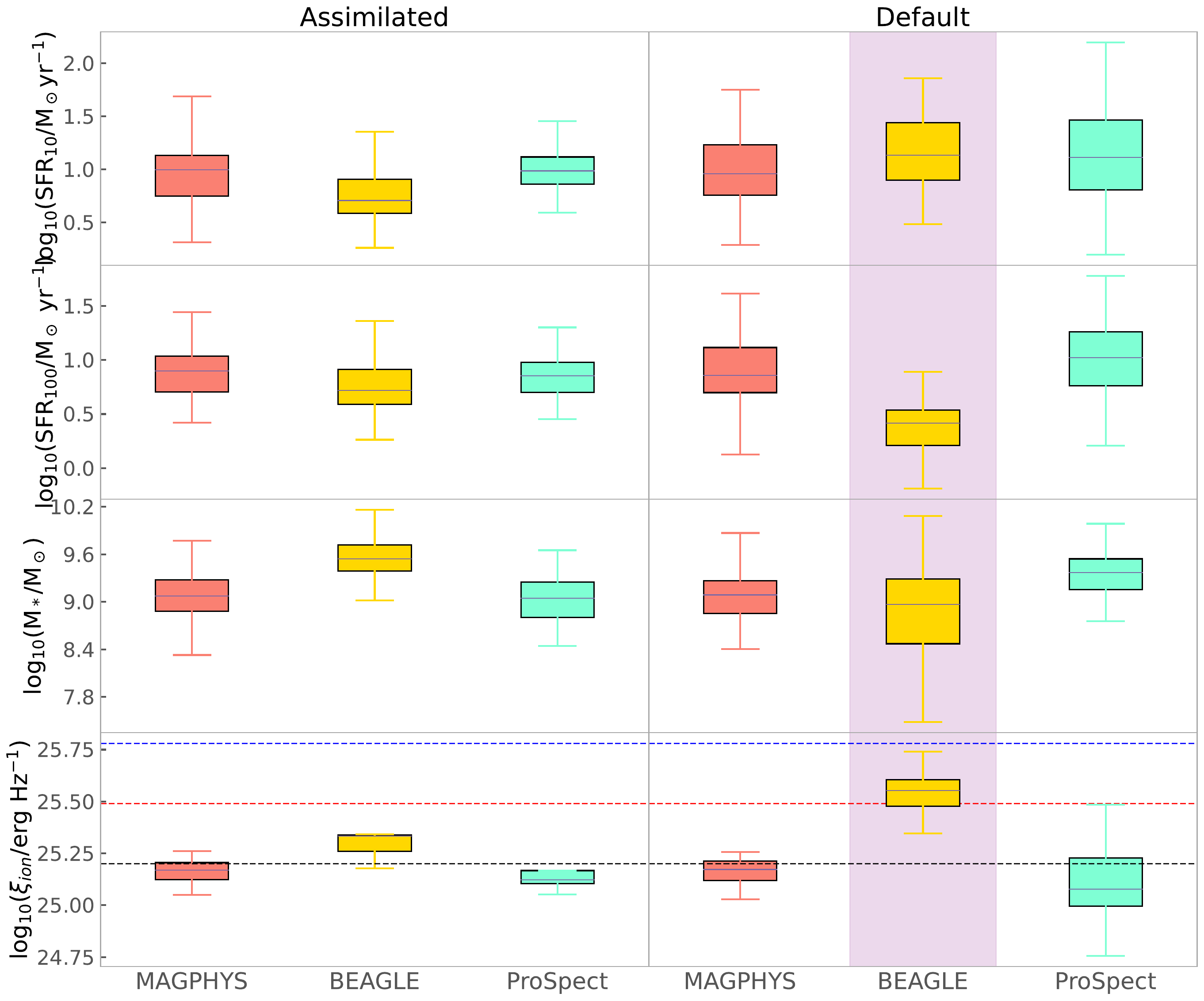}
    \caption{Contrast of the \texttt{MAGPHYS}, \texttt{BEAGLE} and \texttt{ProSpect} SED derived stellar parameter distributions of the JESCO EELG sample with their component models in their most similar (left) vs default (right, see table \ref{tab:Default}) states. The black dashed line reflects the canonical value of 25.2 determined by \cite{Robertson2013} while the red and blue dashed lines respectively reflect the spectroscopic \siion\ median derived using the \texttt{MAGPHYS} and \texttt{BEAGLE} $\rm{L_{UV}(1500)}$ for the subsample with available spectroscopic data. The purple highlighted region represents experiments using the photoionization model. Boxes represent the IQR (75-25 quartiles) with the median value represented by a gray line. Whiskers represent Q1 - 1.5$\times$IQR and Q3 + 1.5$\times$IQR. The default prescriptions of each model create significant variation in the measured parameters for this sample.}
    \label{fig:Default}
\end{figure*}

\subsection{ProSpect}\label{subsec:prospect}
\texttt{ProSpect} \citep{Robotham2020} is a more customizable SED fitting package, including multiple pre-programmed SFH (see Table \ref{tab:big_table} for the ones selected in this study) including non-parametric forms, SPS models, as well as variable priors, dust and emission models (there is no IGM absorption model). By default, \texttt{ProSpect} uses the BC03 stellar population models \citep{Bruzual2003}, \cite{Charlot2000} dust models, \cite{Dale2014} dust emission, \cite{Chabrier2003} IMF, and a truncated skewed Normal star formation history model.

We take advantage of four included SFH models and two SPS models to determine the influence of each assumption on the derived \siion\ and stellar parameters in sections \ref{subsec:SFH} and \ref{subsec:SPS}, respectively. We note that \texttt{ProSpect} is highly customizable, allowing one to implement their own models of SFH or choose the IMF within the SPS model.  

\texttt{ProSpect} also includes an integration with the MAPPINGS photoionization and steady supersonic shock spectral emission code \citep{Dopita1996}. However, the integration of the photoionization code is not self-consistent with the underlying continuum. For the sake of constraining stellar parameters, this has been excluded from this study for this reason. There is also a set of metallicity evolution models however this was not explored in this paper and set to match the {linearly evolving form in \texttt{MAGPHYS} beginning at Z=0.02.} 

\subsection{BEAGLE}\label{subsec:beagle}
\texttt{BEAGLE} (BayEsian Analysis of GaLaxy sEds, \cite{Chevallard2016}, Charlot \& Bruzal in prep) is an SED fitting tool integrated with the CLOUDY \citep{Ferland2017} photoionization model. It allows a range of priors and output parameters, including its own derivation of the \siion\, though it replicates the method used in section \ref{subsec:siionmethod} . By default it uses the BC16 stellar models (Charlot \& Bruzal in prep), \cite{Charlot2000} dust models, a \cite{Chabrier2003} IMF, and a delayed exponentially declining star formation history model with the same functional form as in \texttt{MAGPHYS} but without a burst implementation. While \texttt{BEAGLE} uses a non-bursty delayed exponentially declining star formation history, it optionally includes a constant star formation history (CSFH) portion at the final epoch (default 10 Myr) that could reproduce a recent burst in star formation activity as explored in section \ref{subsec:Bursts}.

We take advantage of the multiple included dust models to test the influence of this choice on the \siion\ in section \ref{subsec:Dust}.  It includes both the classic BC03 \citep{Bruzual2003} and updated BC16 (Charlot \& Bruzal in prep) SPS model with updated isochrones that is explored alongside \texttt{ProSpect}'s available SPS models in Section \ref{subsec:SPS}. The integrated photoionization codes influence is explored in sections \ref{subsec:Bursts} and \ref{subsec:Dust} and used explicitly in section \ref{subsec:spectrophotometry} for the evaluation of what we refer to as the SED+P method (see section \ref{subsec:siionmethod} for an explanation of this method). 

\subsection{Derivation of \siion}\label{subsec:siionmethod}
The production efficiency of Lyman continuum light (\siion) can be derived in a number of ways, reflecting different stages of interaction with the ISM. This paper will explore the model dependence of \siion\, determined through direct photometry in two of these methods: (i) direct integration of the photometric SED flux below 912\AA\ \citep{Wilkins2016a} and (ii) SED+P - using photoionization modeled emission line flux from the photometric SED (see \citealt{Bouwens2015} for a similar approach using excess flux). These will also be compared to a spectroscopically derived \siion\, values in section \ref{subsec:spectroscopy}.

\subsubsection{Photometric Method 1. Integration}
The direct photometric SED integration method determines the intrinsic LyC production using the dust-corrected modeled SED of a galaxy as follows:
\begin{equation}
    \xi_{ion}/({\rm Hz}\  {\rm erg}^{-1}) = \int^{c/912\AA}_{c/-\infty} L_\nu ({\rm h}\nu)^{-1}/L_\nu(1500\AA) {\rm d}\nu,
\end{equation}
where $c$ is the speed of light in \AA/s and $h$ is the Planck constant in $\mathrm {erg\ Hz^{-1}}$. This method integrates the total unattenuated LyC luminosity and normalizes it by the non-ionizing, unattenuated UV luminosity density at 1500\AA. For the \siion\, derived from the \texttt{MAGPHYS} and \texttt{ProSpect} fitting codes, we perform the integration as described. For the \texttt{BEAGLE} code we use the internal \texttt{\siion\_unatt} parameter which uses a similar approach \citep{Chevallard2017}. 

This is the method we primarily explore in this paper in terms of its model dependence.

\subsubsection{SED+P Method \& Spectroscopic Method}
The SED+P method requires an SED fitting code with an integrated photoionization model \citep{Dopita1996,Ferland2017}. The modeled H$\beta$ or H$\alpha$ emission lines are treated as it would be in the pure spectroscopic method: 
\begin{equation}
   \xi_{ion} = \frac{N(H^0)}{L_{UV}\times c_{rec}}, 
\end{equation}
where N(H$^0$) = 7.28$\times$10$^{{\rm 11}}L(H\alpha)$  \citep{Leitherer1995} indicating \fesc=0 with the zero superscript \citep{Simmonds2023} and $c_\mathrm{rec} = 2.86$ is the case B recombination constant if using $L(H\beta)$ \citep{Osterbrock} at T=10$^4$ K and n$_e$=100 cm$^{-3}$. The spectroscopic and SED+P derivations use the interaction of LyC photons with nebular gas in the ISM that produces hydrogen recombination lines such as H$\alpha$ and H$\beta$. The total flux of hydrogen recombination lines can then be reverse engineered to reflect the initial production of ionizing photons assuming none of them escape. There is also a caveat for the direct LyC absorption by dust which will be missed by nebular line methods \citep{Smith2022}, however, as our sample not dusty (\cite{Jaiswar2024}, $\rm{A_V}$ $0.25^{+0.12}_{-0.14}$) we do not explore this possibility.    

For clarity, each method is explored separately, with the direct integration method explored in sections \ref{subsec:Default}-\ref{subsec:Dust} for the model dependence and the SED+P method explored in section \ref{subsec:spectrophotometry} for its proximity to the true spectroscopic value, where the spectroscopic method is also explored.

One advantage of the SED+P method is that modeled emission lines consider modeled dust absorption, hence they reflect the intrinsic emission line. Mixing SED derived stellar attenuation with nebular attenuation introduces an additional uncertainty from the wide scatter of this relation \citep{Reddy_2020}. \texttt{BEAGLE} calculates an observed and emitted (intrinsic, before attenuation) emission line value which in our sample differ by less than 0.5\%. To dust correct real spectra requires coverage of both H$\alpha$ and H$\beta$ which is not available for our KMOS sample, and reduces our prism sample further, hence we do not dust correct the spectroscopic values.         

\section{Results}\label{sec:results}

\subsection{Default vs Assimilated model}\label{subsec:Default}
While each SED code can be made to approximate one another , this is often impractical to execute and may undermine the intended purpose of the chosen code. We therefore explore the differences in the derived parameters from each code in the default state as well as their most similar state (henceforth ``assimilated" state) as the default likely reflects general usage while assimilation reflects consistency. 

Here and throughout the paper we focus on 4 key stellar parameters: Star formation rate over 10 and 100 Myr timescales (SFR$_{10}$ and SFR$_{100}$), stellar mass (M$_*$) and the ionizing photon production efficiency (\siion). We highlight the limitations in measuring SFR$_{10}$ in \texttt{ProSpect} and \texttt{MAGPHYS}. Neither of these codes self-consistently incorporate a photoionization model while fitting the SED, which is necessary to constrain short timescale star formation with photometry alone. However, as we are exploring the parameterization itself rather than the ``true" value, we include the SFR$_{10}$ measurements in the analysis for all codes.

\begin{table}
    \centering
    \begin{tabular}{c|c|c|c}
         & MAGPHYS & ProSpect & BEAGLE \\ \hline
         SFH & d$\tau$ & trunc skew Normal & d$\tau$+CSFH \\
         bursts & yes & no & no\\
         SPS & BC03 & BC03 & BC16 \\
         dust &CF00 & CF00 & CF00 \\
         metal evol & linear & variable & linear \\
         IGM & Madau95 & none & Inoue14 \\
         neb emission &none &MAPPINGS& CLOUDY \\
         $\chi^2$& 21.6& 17.3&8.3\\
    \end{tabular}
    \caption[Default models and median residuals]{True default models used by each SED code and their median residual $\chi^2$ (not reduced), where a lower value represents a stronger agreement between the photometry and SED fit. More details can be found in section \ref{sec:methods}. IGM absorption was removed from \texttt{MAGPHYS} and nebular emission was removed from \texttt{ProSpect} for all experiments, including \ref{subsec:Default} due to their implementation or in order to access the unimpeded LyC flux. {The metallicity evolution in \texttt{ProSpect} was also set to be linear in all experiments as in \texttt{MAGPHYS} though this was not explored in this paper.}}
    \label{tab:Default}
\end{table}

\begin{figure*}
    \centering
\begin{minipage}[t]{0.50\textwidth}
    \centering
    \includegraphics[width=\textwidth]{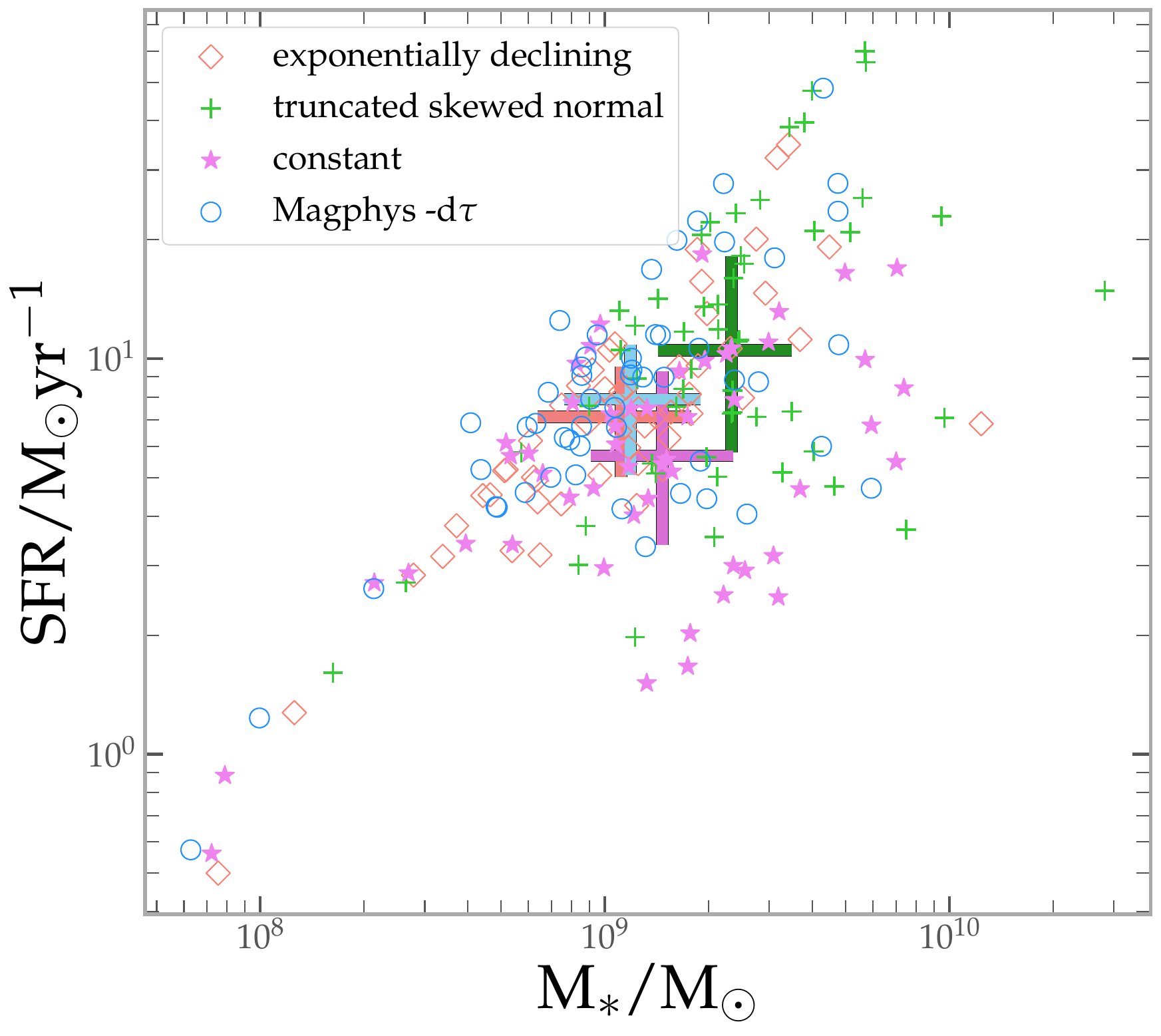}
\end{minipage}\hfill \begin{minipage}[t]{0.30\textwidth}
    \centering
    \includegraphics[width=\textwidth]{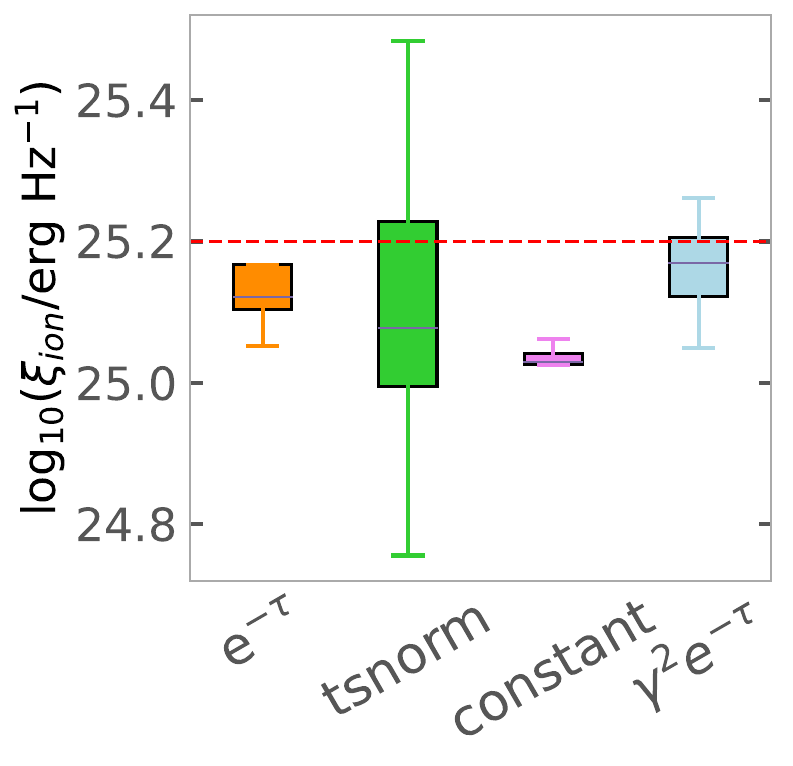}
\end{minipage}\hfill \begin{minipage}[t]{0.18\textwidth}
    \centering
    \includegraphics[width=\textwidth]{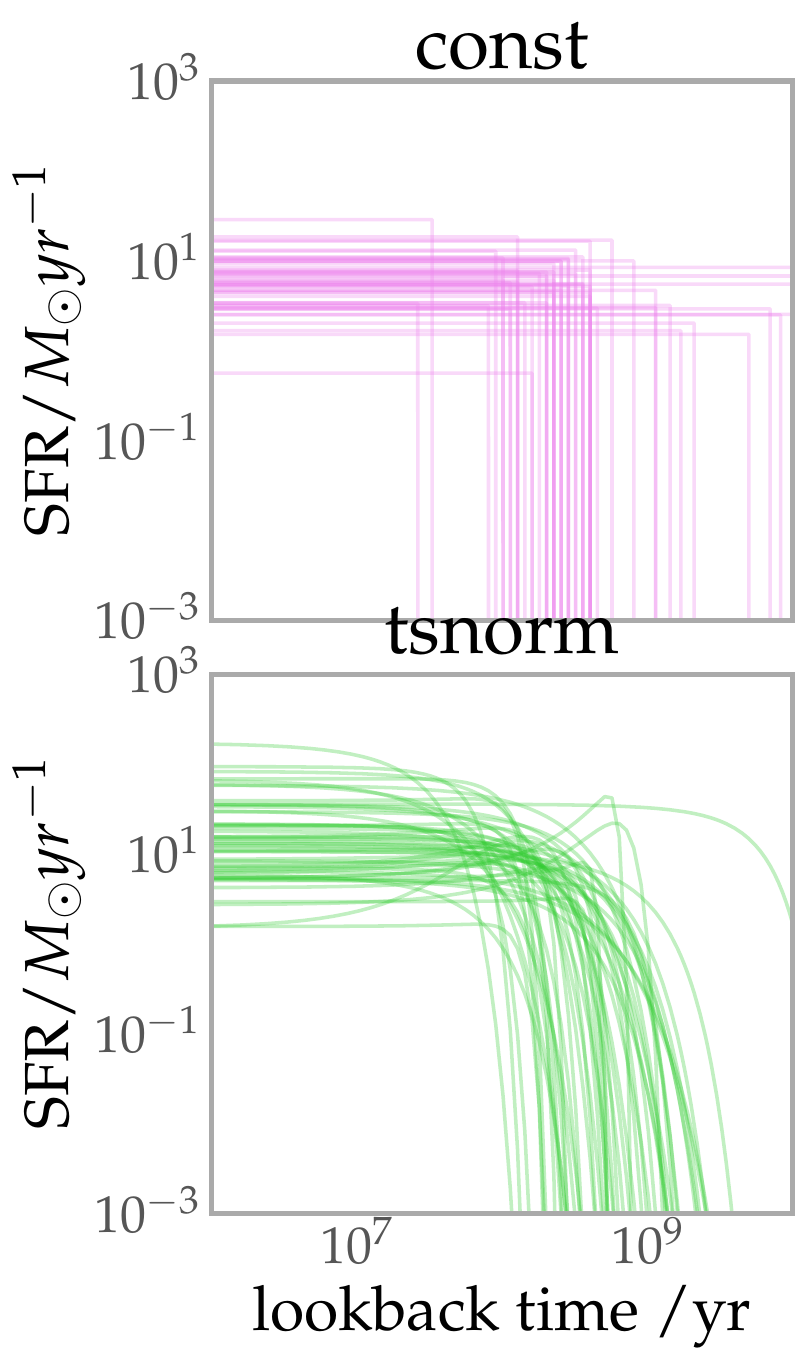}
\end{minipage}
    \caption{(Left) Main sequence comparison using four different star formation history models from \texttt{MAGPHYS} and \texttt{ProSpect} of the JESCO EELG sample (using SFR$_{100}/$M$_*$). Coloured crosses reflect the 25-75\% percentiles of the sample matching their color. (Middle) ionizing photon production efficiency of the full sample using the different star formation histories, where the red dotted line represents the canonical value of \siion\ \citep{Robertson2013}. (Right) Comparison of two SFHs fit for the constant (pink) and truncated skewed Normal (green) models.}
    \label{fig:SFH}
\end{figure*}

The codes were assimilated by modifying the more malleable \texttt{BEAGLE} and \texttt{ProSpect} models toward a central point resembling the more rigid \texttt{MAGPHYS} code. We did, however, remove the inclusion of stochastic bursts in \texttt{MAGPHYS} as replicating this exact method is not possible on the user end of the other models. 

\begin{figure*}
\centering
    \includegraphics[width=0.75\linewidth]{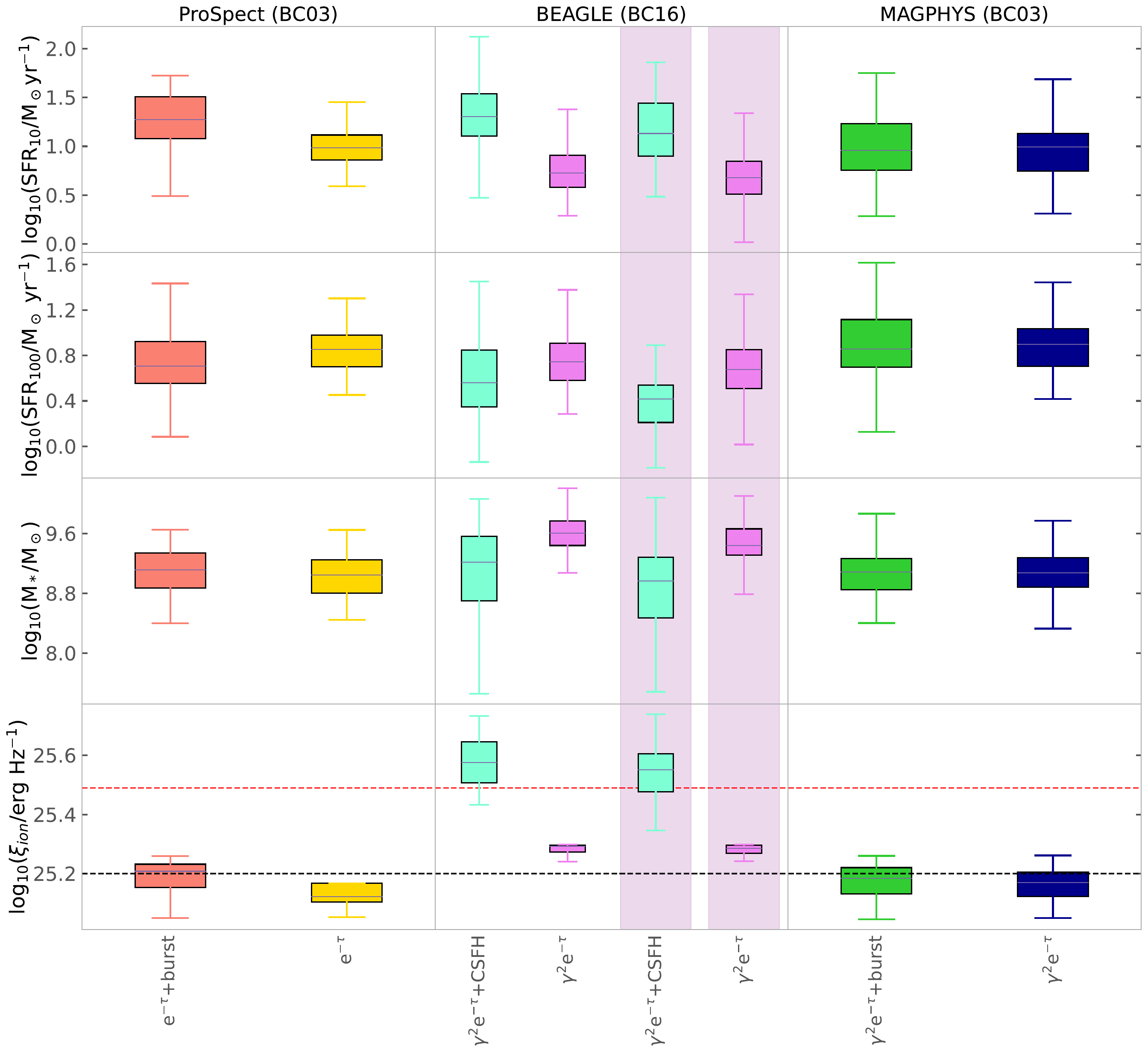}
\centering
    \includegraphics[width=0.45\linewidth]{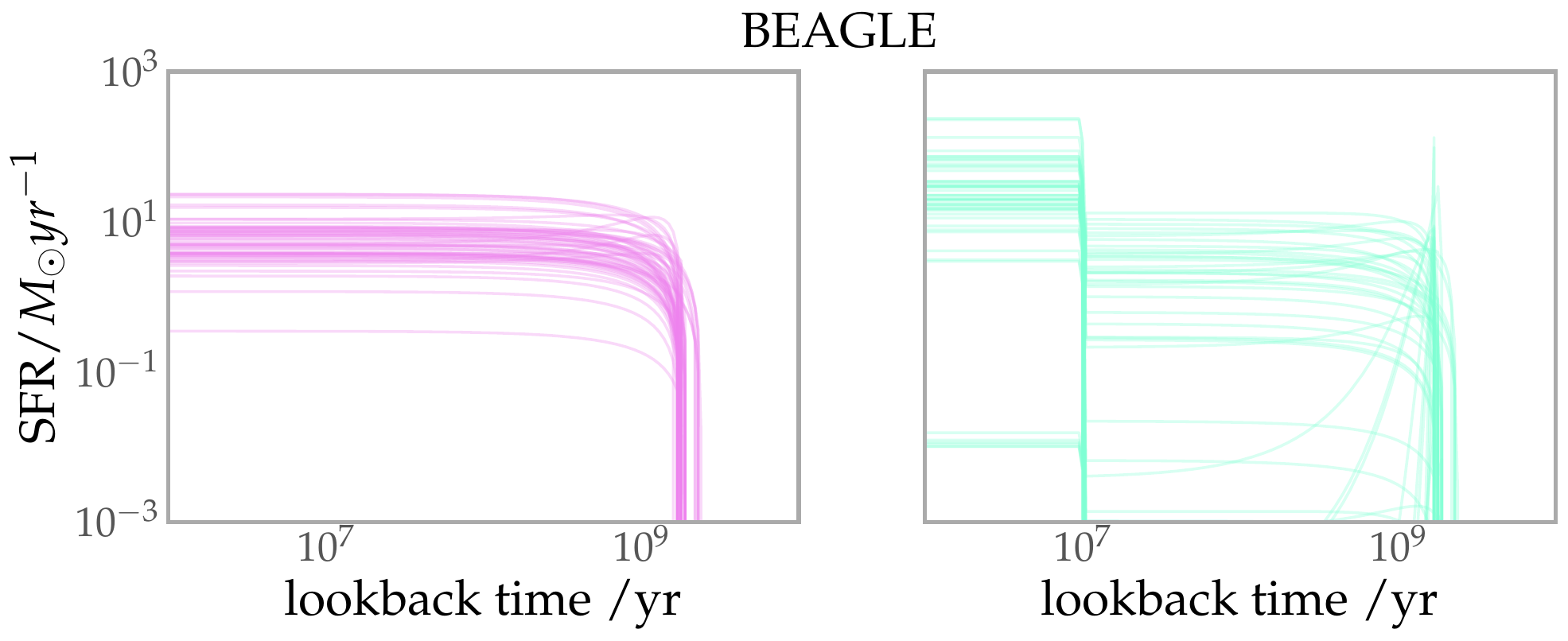}\includegraphics[width=0.45\linewidth]{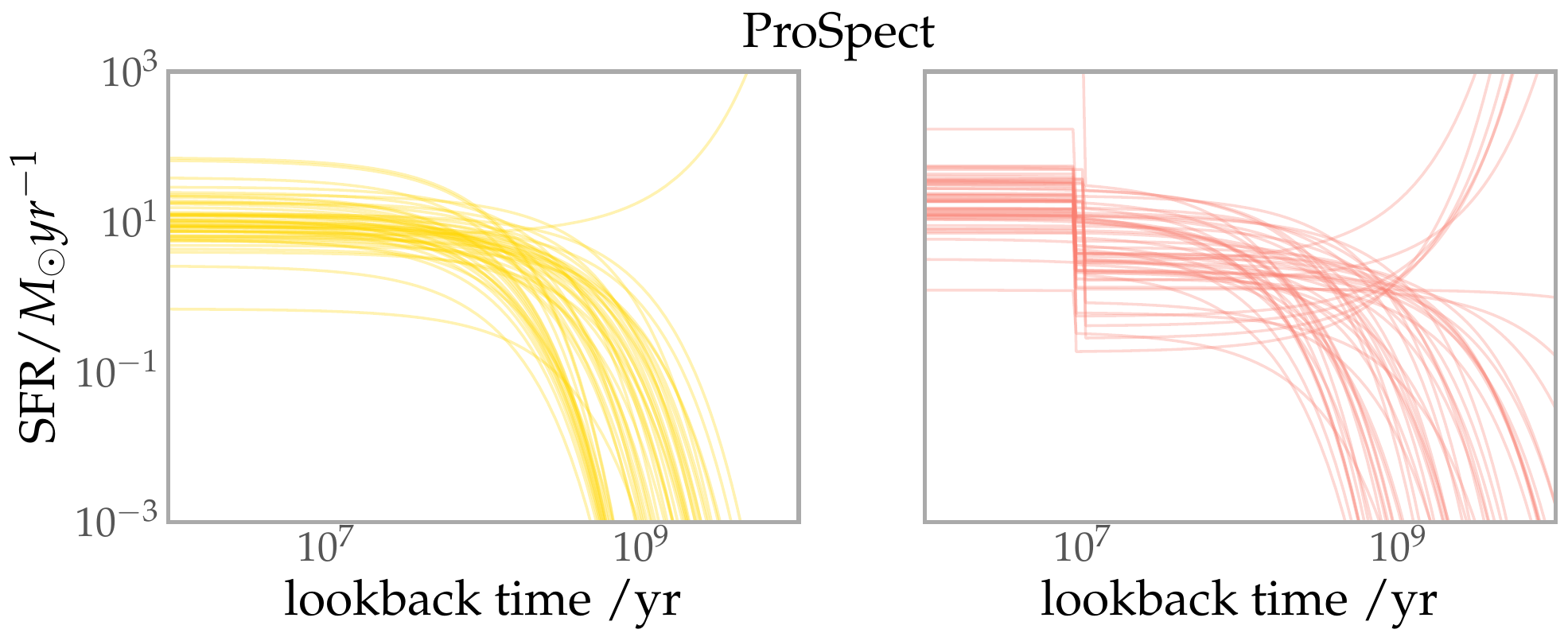}\par\vspace{1em}
    \includegraphics[width=0.45\linewidth]{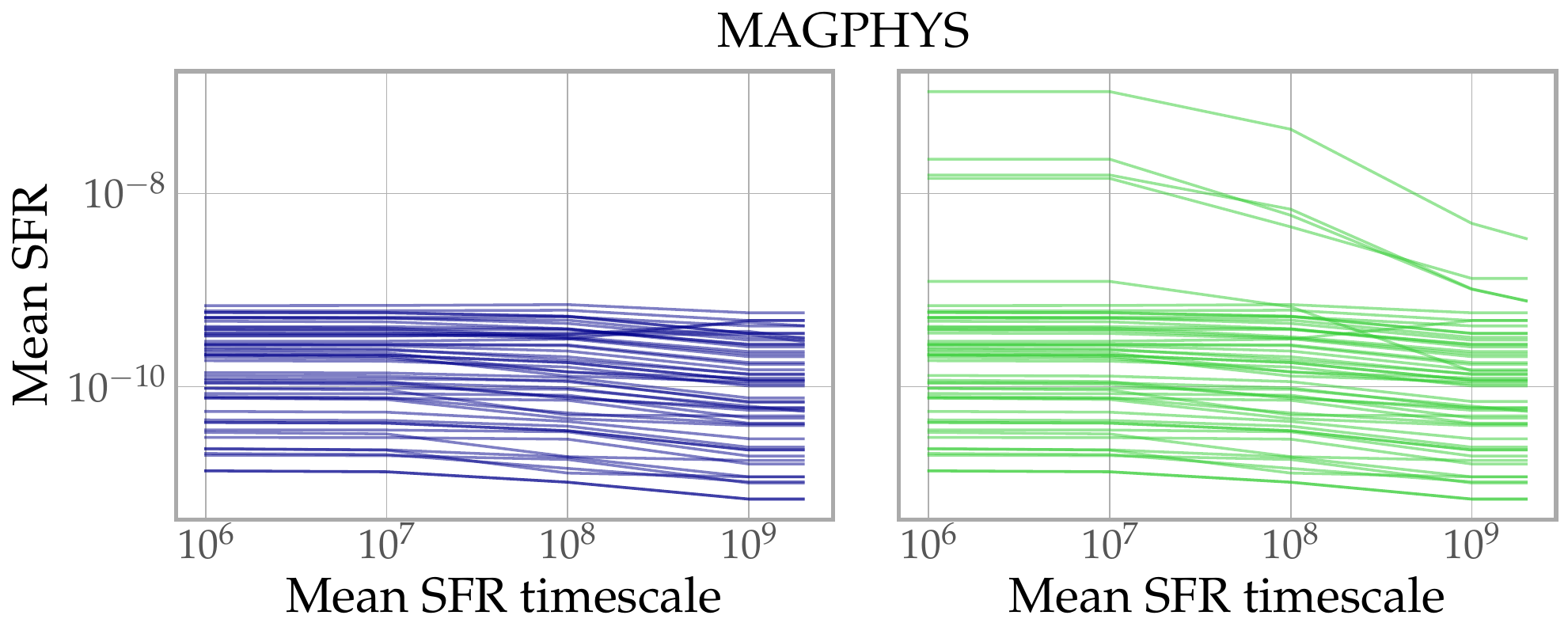}
\caption{ (Top) Derived stellar parameters with and without bursts for each SED fitting code. $e^{-\tau}$ refers to the exponentially declining SFH while $\gamma^2e^{-\tau}$ refers to the delayed exponentially declining SFH. The CSFH portion of the \protect\texttt{BEAGLE} SFH is treated as a burst for this analysis and is contrasted with the burst implementation of the other models. Purple highlighted regions represent models including the photoionization model CLOUDY \citep{Ferland2017}. Black and red dotted lines represent the canonical \citep{Robertson2013} and the median spectroscopic value derived using the \protect\texttt{MAGPHYS} L$_{UV}$ respectively. Boxes represent the IQR (75-25 quartiles) with the median value represented by a grey line. Whiskers represent Q1 - 1.5$\times$IQR and Q3 + 1.5$\times$IQR. We find that the constant SFH portion of the \texttt{BEAGLE} model for our bursty population consistently shifts the \siion\, to larger values compared to our other models. (Bottom) Corresponding SFH for BEAGLE (Left) ProSpect (Right) and average SFH for MAGPHYS (Middle) shown with colors corresponding to the models used above.}
    \label{fig:Burst}
\end{figure*}

The \texttt{BEAGLE} SFH by default, includes a CSFH component as a free parameter in the last 10$^6$ years, replacing the d$\tau$ SFH (see Figure \ref{fig:Burst}). This alters the final stellar population from which \siion\, is derived. Accordingly, so this final portion was removed in the assimilated version. The default BC16 SPS model with nebular emission fitting was also changed to the BC03 SPS model that does not include nebular modeling.

\texttt{ProSpect} uses the more flexible truncated skewed Normal distribution for its SFH by default, enabling it to form a rising, decaying or Normal SFH by shifting the Normal peak beyond the galaxy's lifespan. This was changed to the simpler exponentially declining SFH model in the assimilated experiment. We select the CF00 dust model and the BC03 SPS model. Figure \ref{fig:Default} shows a comparison of key properties between the default and assimilated states. The full model list can be found in Table \ref{tab:Default}. 

Removing the burst in the assimilated \texttt{MAGPHYS} model has a minor impact to all measured parameters for our sample. The way bursts are implemented in \texttt{MAGPHYS} is such that the probability of a burst occurring in the last 2 Gyr is set to 75\%; these may last between 30-300Myr and the stellar mass formed in this time can be anywhere from 0.1 to 100 times the value of the underlying SFR \citep{DaCunha2015}. As this timescale matches that of the total universe age at z$\sim$3, the bursts tend to occur at the formation epoch, and only very few ($\sim$9\%) undergo a subsequent burst. Removing bursts, therefore, only significantly changes the final SFR for the few galaxies experiencing secondary bursts. 

In \texttt{BEAGLE} however, the default and assimilated models differ significantly with changes to the CSFH free parameter, photoionization modeling and SPS model (Figure \ref{fig:Default}). The SFR$_{10}$ is reduced by more than half (\SF=0.43$\sim$dex) while the median SFR$_{100}$ doubles. The median stellar mass value is larger by $\sim$30\% in the assimilated \texttt{BEAGLE} model and occupies a stricter range when contrasted to the default setting results. Notably, the median assimilated \siion\, is 0.3 dex below its default value with no overlap in the inter-quartile range (IQR).

Comparing the default and assimilated versions of \texttt{ProSpect}, we find the median \SF = 0.1 dex each, with their IQRs reflecting $\sim$25\% and $\sim$35\% of their original respective scatters. In contrast to \texttt{BEAGLE}, this represents a significant reduction in both the short and long timescale SFR median and scatter. The median stellar mass drops by $\sim$0.3 dex while maintaining a similar scatter, but the \siion\, IQR narrows to a third of its original range.    

A comparison of the three models in their default states highlights contrasting behaviors. The SFR$_{10}$ IQR of \texttt{BEAGLE} and \texttt{ProSpect} overlap entirely, with the median differing by \SF=0.02 dex. In contrast, \texttt{MAGPHYS} exhibits a lower median log$_{10}$(SFR$_{10}$) by approximately \SF = 0.15 dex and a 40\% smaller IQR. For the longer timescale SFR$_{100}$, \texttt{MAGPHYS} and \texttt{ProSpect} are much larger, showing medians over three times higher and IQRs approximately five times wider than \texttt{BEAGLE}. \texttt{MAGPHYS} allows greater scatter in SFR$_{100}$, which it reasonably constrains, while \texttt{BEAGLE} imposes stricter limits on SFR$_{100}$ to compensate for its more precise handling of SFR$_{10}$.

\texttt{ProSpect}'s default behavior diverges mainly due to its SFH formalism , as metallicity evolution and MAPPINGS photoionization integration are not considered here. The flexibility of the truncated skewed Normal model is reflected in the larger scatter observed in both SFR$_{10}$ and SFR$_{100}$. \texttt{ProSpect} estimates the median stellar mass of these EELGs to be about $\Delta\log$(M/M$_\odot$)=0.4 dex higher than both \texttt{MAGPHYS} and \texttt{BEAGLE}, with a scatter comparable to \texttt{MAGPHYS}. \texttt{BEAGLE}, however, shows a broader IQR and lower mass limits, likely due to its constrained SFR$_{100}$, though other factors may contribute. Finally, \texttt{BEAGLE} calculates a median LyC production efficiency over 0.3 dex higher than \texttt{MAGPHYS} and the canonical value from \citep{Robertson2013}, and more than 0.4 dex higher than \texttt{ProSpect}.

Many of the noted discrepancies are smoothed out with the assimilation of these models. \texttt{MAGPHYS} and \texttt{ProSpect} calculate similar median SFR$_{10}$ and SFR$_{100}$ which are around 0.24 dex and 0.2 dex larger than \texttt{BEAGLE}'s respectively. With assimilation, each model favors stricter constraints on the shorter timescale SFR relative to the SFR$_{100}$. \texttt{BEAGLE} now estimates a larger mass compared to \texttt{MAGPHYS} and \texttt{ProSpect} by 0.5 dex and has a similar amount of scatter in this sample. The \siion\, median and IQR is also found to match more closely between the three models, yet, there is still a discrepancy with the \texttt{BEAGLE} median being $\sim$0.15 dex higher than \texttt{MAGPHYS}, suggesting some unaccounted for source of extra ionizing budget.     

Of the 6 model combinations in Figure \ref{fig:Default}, we find that only the default \texttt{BEAGLE} model lies within the spectroscopic median value derived using the \texttt{MAGPHYS} $\mathrm{L_{UV}(1500)}$\AA\, (red line in Figure \ref{fig:Default}) and \texttt{BEAGLE} $\mathrm{L_{UV}(1500)}$\AA\, (blue line) while the rest appear similar to the canonical \cite{Robertson2013} value. The default configuration of BEAGLE therefore significantly increases the measured \siion\, which would explain some of the observed discrepancy between the Tang23 sample in \cite{Jaiswar2024} and the original \cite{Tang2023} study of the same sources. The significance of the width of each parameter range and how they depend on models cannot be overstated, as this is a uniform sample which preemptively limits the dynamical range. {The physical interpretation for this difference is explored in sections \ref{subsec:Bursts} and \ref{subsec:spectrophotometry}.}    

\begin{figure*}
    \centering
    \includegraphics[width=0.75\linewidth]{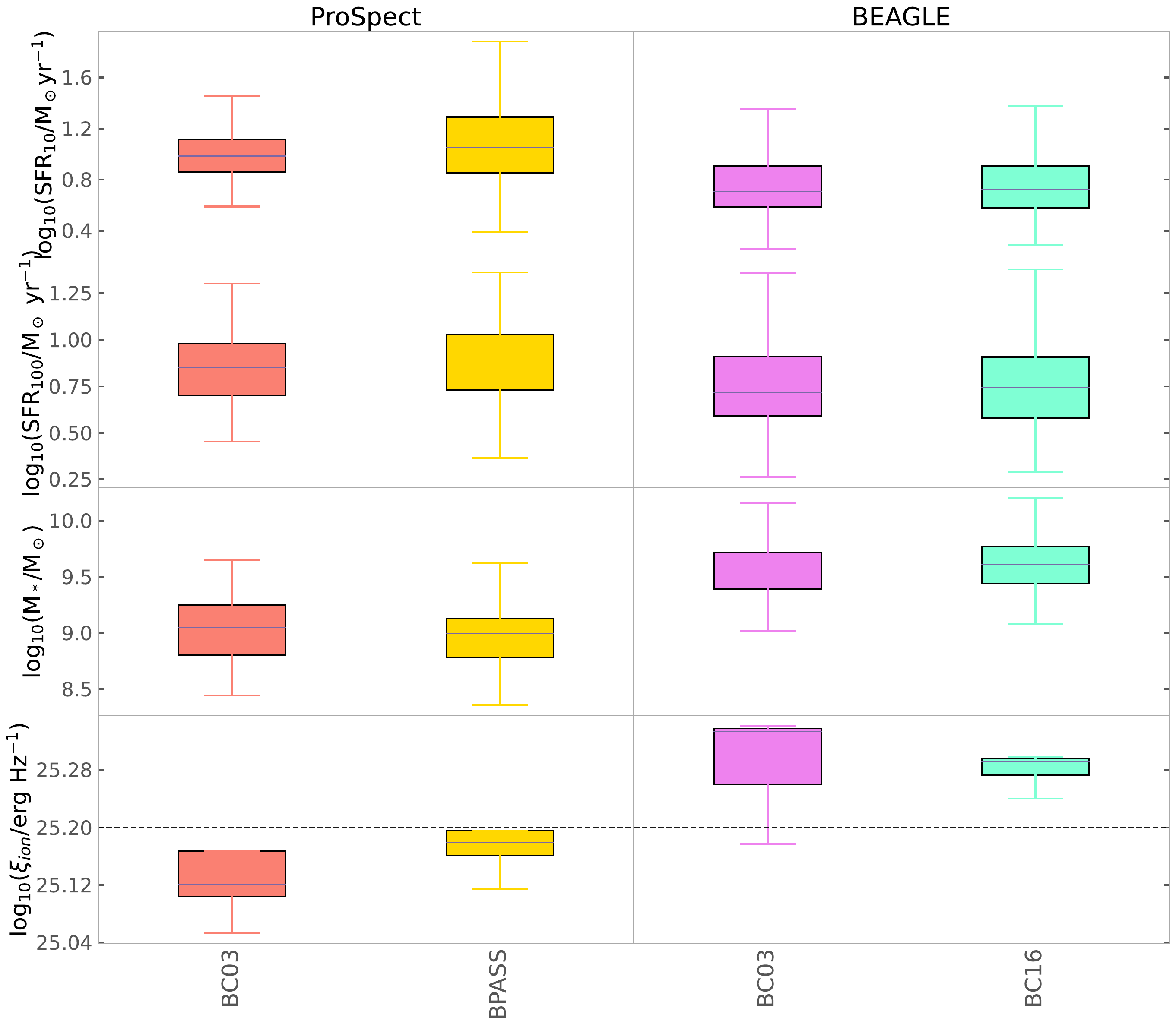}
    \caption{Comparison of SPS models without included photoionization modeling and no bursts. Black line represents the 25.2 canonical \siion\ value \citep{Robertson2013} Boxes represent the IQR (75-25 quartiles) with the median value represented by a gray line. Whiskers represent Q1 - 1.5$\times$IQR and Q3 + 1.5$\times$IQR. We find that the BPASS model estimates a consistently higher value of \siion\, relative to the BC03 model while the BC16 model is consistently below the median BC03 in their respective fitting codes.}
    \label{fig:SPS}
\end{figure*}

\subsection{Dependence of \siion\ on SFH }\label{subsec:SFH}
The choice of SFH model should balance complexity with expectations of the stellar population. A constant SFH is unrealistic for most galaxies; however, at extremely high redshift, the relatively short lifespan of a galaxy may not reflect additional complexity. A delayed exponentially declining model is the most commonly seen default SFH model \citep{DaCunha2008,DaCunha2015,Robotham2020}, capable of simultaneously creating an initially rising SFH reflecting galaxy formation, and the eventual decline as it ages. A reasonable expectation is that as the choice of model determines the relative proportion of stars at different ages, this would then be reflected in the \siion\ as short lived massive stars are its primary drivers. While implementation of bursts is relevant to the SFH model, we find that this warrants its own section (see section \ref{subsec:Bursts}). 

We use \texttt{ProSpect} to explore three different SFH models- exponentially declining, truncated skewed Normal, constant- and compare them to \texttt{MAGPHYS}'s delayed exponentially declining model. For details on their parameterization refer to \cite{Robotham2020} and the \texttt{ProSpect} documentation, and \cite{DaCunha2015} for the delayed exponentially declining model in \texttt{MAGPHYS}. Put simply, the delayed exponentially declining and exponentially declining models differ by an additional $\gamma^2$ multiplicative factor, where $\gamma$ represents the inverse of the star formation timescale. The truncated skewed Normal model is a Normal distribution with a skewing factor to allow greater variability in SFH shapes and a truncation of the SFH at its formation epoch. The constant SFH assumes a single SFR value sustained over the galaxy's formation time. 

The closeness of stellar mass and star formation rates (Figure \ref{fig:SFH}) for the two exponentially declining models reveals the overall similarity between them, with \texttt{MAGPHYS} delayed exponentially declining calculating a median SFR$_{100}$ 0.05 dex higher, and an almost identical SFR$_{10}$ and stellar mass. The \texttt{MAGPHYS} model overlaps with the \cite{Robertson2013} canonical value (log(\siion/[erg Hz$^{-1}$])=25.2) at its upper quartile. The exponentially declining model derives a slightly lower (log(\siion/[erg Hz$^{-1}$])=25.12) median \siion\ corresponding to the 25th percentile of the delayed exponentially declining model with its 75th percentile equaling the median delayed exponentially declining \siion. 

The truncated skewed Normal distribution, in contrast, estimates a higher median stellar mass (by 0.3 dex) and SFR over both timescales (\SF = 0.13 dex SFR$_{10}$ and \SF = 0.17 dex SFR$_{100}$ respectively) with a significantly larger scatter in SFR. The wider range is also reflected in the derived \siion\, which has the widest range of these experiments (IQR $\Delta\log_{10}/\rm{erg~ Hz^{-1}}$ = 0.23 dex). While the truncated skewed Normal distribution has a functional form, it has been shown to predict non-functional (stepwise or so-called non-parametric) stellar mass distributions, and this mass offset compared to \texttt{MAGPHYS} is documented \citep{Thorne2021}. 

The constant SFH determines a slightly higher median stellar mass ($\Delta\log$(\M=0.1 dex) but lower SFR$_{100}$ (\SF=0.15 dex) than the delayed exponentially declining model, and has the most narrow \siion\ range of any experiment. Based on individual fit quality, we suspect that the increased mass and SFR range is achieved by linearly offsetting a model that closely resembles the observed flux densities, though with limited capacity to fit the full photometry. This is aided somewhat by varying the formation age, which was implemented for this parametrization only, though the SED shape remains relatively the same. Offsetting a similar model also results in similar relative integrated LyC and $\mathrm{L_{UV}(1500)}$\AA\,, which yields a narrow range of \siion.  

There is a debate as to whether parametric SFH models should be used in studies exploring diverse galaxy samples, as a significant portion may be poorly modeled by the rigidity of the assumed functional form \citep{Leja2019}. However nonparametric models suffer from limited time resolution information. Here, the \siion\, ranges appear to correlate with greater flexibility in the SFH parameterization, with more degrees of freedom resulting in a wider range of \siion. 

\subsection{Dependence of \siion\ on Bursts}\label{subsec:Bursts}
A bursty SFH is traditionally defined by strong fluctuations in SFR over short periods. This simplistic definition is complicated in practice by its unique implementation in different SED codes and their intentions. The high\_z version of \texttt{MAGPHYS} \citep{DaCunha2015}  is designed for $z>1$ galaxies and follows a similar burst implementation to the original version \citep{DaCunha2008}, allowing multiple (or 0) bursts of varying duration and SFR to occur throughout its history with the goal of best fitting the resultant dust and stellar properties for a given flux density profile. This works best at low redshifts as the formation history extends far back enough to allow multiple bursts to occur and capture the true stochastic nature of star formation. 

\texttt{BEAGLE} on the other hand does not include traditional bursts of star formation, favoring a smooth delayed exponentially declining SFH. However, the optional inclusion of a portion of constant SFH (CSFH) as a free parameter at a set epoch may act as a burst of star formation if it determines a rate above that of the preceding continuum. Its a mathematical convenience with the primary objective of self-consistently accounting for nebular emission without forcing the preceding history to accommodate it. By default, this is set to occur 10 Myr prior to the current epoch, intended to provide H$\alpha$ derived SFR$_{10}$ constraints.  

\texttt{ProSpect} SFHs may also optionally include bursts, set by default to occur 100 Myr in lookback time. This burst overlays the existing SFH with a single region of increased SFR, with the intention of recreating a bursty history within the timescale limitations of UV-NIR photometric constraints. We have opted to reduce the burst to occur at 10Myr in lookback time to make comparisons to \texttt{BEAGLE}s method, though we note this timescale is not properly constrained without H$\alpha$ emission (the stellar template resolution is sufficient for this choice). Unlike \texttt{MAGPHYS}, both \texttt{BEAGLE} and \texttt{ProSpect} force this singular event at the specified epoch.

In \texttt{MAGPHYS}, our results  (see Figure \ref{fig:Burst}) indicate that the inclusion of bursts has a limited influence on the derived stellar parameters for our sample, however, it has a complex relationship in the other models. As bursts are allowed anywhere in the last 2 Gyr, and the formation and evolutionary history of these galaxies is within this range, most \texttt{MAGPHYS} galaxies experience their `burst' at their inception with only 9\% experiencing subsequent bursts in the current epoch. The removal of bursts for this redshift population therefore has only a minor effect on the derived parameters.

In absence of a burst, the \texttt{MAGPHYS} and \texttt{ProSpect} models appear to produce very similar SFR$_{10}$(\SF = 0.01 dex), however, inclusion of bursts appears to essentially double \texttt{ProSpect}s SFR$_{10}$(\SF=0.31 dex relative to \texttt{MAGPHYS}). Burst inclusion also slightly reduces the median \texttt{ProSpect} SFR$_{100}$ by $\sim$30\%. These phenomena are a consequence of the short timescale choice for this burst and the fact that bursts occur in the entire sample. The reduced SFR$_{100}$ is due to the continuum before 10 Myr reducing to accommodate the additional mass produced in the burst. We see that overall this has a minor impact on the calculated median stellar mass $\Delta\log$(M/M$_\odot$)=0.06 dex. We do observe an uptick in calculated \siion\, of 0.1 dex, which is expected with the increased massive stellar population's added contribution to the LyC photon production.    

The integration of \texttt{BEAGLE}s photoionization model creates an additional parameter requiring consideration when making the direct comparison to the other models. This integration is specific to the BC16 \citep{Gutkin2016} SPS model, so a further consideration must be made for the significance of the chosen SPS model. However, for the sake of clarity this discussion has been left to section \ref{subsec:SPS}. The full matrix of combinations is available in Table ~\ref{tab:big_table}.

We find as with \texttt{ProSpect} that the inclusion of the CSFH portion in our sample significantly increases the SFR$_{10}$ relative to the CSFH free model, both with nebular emission considered (\SF=0.45 dex) and in absence of it (\SF=0.57 dex). The burst and no burst values also correlate with the corresponding \texttt{ProSpect} values in absence of the photoionization model. This not only suggests very similar behavior between the \texttt{ProSpect} burst and this free CSFH parameter but also backing the identity of these galaxies as undergoing current bursts \citep{Cohn2018}.

Inclusion of photoionization reduces the median SFR$_{10}$ by $\sim$30\% which is likely due to the lower mass formed. Bursts also reduce the median SFR$_{100}$ by $\sim$35\% in absence of photoionization, similar to the $\sim$30\% experienced in \texttt{ProSpect}, and by $\sim$45\% for the photoionized model. Notably, \texttt{BEAGLE} calculates a lower median stellar mass with bursts in both cases ($\Delta\log$(M/M$_\odot$)=0.47 dex  with photoionization and $\Delta\log$(M/M$_\odot$)=0.39 dex without). While these masses overlap better with the \texttt{MAGPHYS} and \texttt{ProSpect} calculated masses the increased range indicates how dependent this final mass determination is on the free CSFH parameter. Furthermore, the models including photoionization calculate a lower stellar mass than those without. This is likely due to the presence of bright nebular emission lines in the derived flux densities of our EELG type galaxies which can be seen also in Figure \ref{fig:Dust} by comparing the continuum of models derived in the presence/absence of the photoionization model. As described in \cite{Jaiswar2024}, these contaminate the containing filter which calculates an increased flux density, thereby increasing the SED continuum and increasing the calculated mass. 

Finally, the \siion\, is found to vary substantially with this included burst, by $\sim$0.3 dex in both photoionized and standard models. This is significantly above the calculated \siion\, in either \texttt{MAGPHYS} or \texttt{ProSpect} suggesting that this free CSFH parameter is a strong determining factor. {Mechanically, \texttt{BEAGLE} appears to calculate both a higher $\mathrm{L_{UV}(1500)}$ (see section \ref{subsec:spectrophotometry}) and higher LyC than the other models (that more than compensates for the higher $\mathrm{L_{UV}(1500)}$). Combined with the short and long term SFR, the physical interpretation is that the excess of massive stars created by the recent burst contribute significantly to the estimated ionizing and non-ionizing UV, and thus raise \siion. We also find that \texttt{BEAGLE} calculates a median $\mathrm{1.4^{2.2}_{1.0}\times}$ larger $\mathrm{L_{UV}(1500)}$ and $\mathrm{2.7^{5.1}_{1.3}}\times $ larger \siion\, when including a burst}. Taken together, these results confirm the sensitivity of BEAGLE's outputs to the inclusion and timing of the CSFH.

\subsection{Dependence of \siion\ on SPS model}\label{subsec:SPS}

Stellar Population Synthesis (SPS) models estimate the relative populations of stellar subtypes from an object's observed luminosity \citep{Eldridge2017}, combining an IMF, stellar isochrones, and fixed stellar spectra. Inside SPS codes, simple stellar populations are defined across ranges of age, metallicity, $T_\mathrm{eff}$, and $L_\mathrm{bol}$ \citep{Conroy2013}. Existing SPS models target specific stellar populations or phenomena \citep{Conroy2009}.

The BC03 model \citep{Bruzual2003}  uses STELIB spectra \citep{LeBorgne2003} and considers thermal AGB pulses, which increases mass loss, chemical synthesis and luminosity \citep{Pastorelli2019}. The BC16 model incorporates updated stellar evolutionary tracks \citep{Marigo2013}, a pre-main sequence phase \citep{Bressan2012}, and improvements to the equation of state and nuclear reaction rates, using the updated MILES spectra \citep{Sanchez-Blazquez2006}  instead of the more limited STELIB library (Charlot \& Bruzual in prep). The BPASS model \citep{Eldridge2017} emphasizes interacting binaries. These are common among massive stars and can significantly alter their evolution \citep{Sana2012,Sana2014} by extending their lifespans \citep{Eldridge2008} and driving the evolution of their companions. Modern SED-fitting tools like BayEsian Analysis of GaLaxy sEds (\texttt{BEAGLE}) integrate the photoionization code CLOUDY, requiring high-resolution stellar population templates to fit nebular emission \citep{Gutkin2016}. 

Here, we compare the three SPS models (Figure \ref{fig:SPS}) as implemented in the \texttt{ProSpect} and \texttt{BEAGLE} codes. However, fair comparisons can only be made within the full framework of each code. Even when using the same BC03 model without incorporating burst prescriptions or photoionization, \texttt{BEAGLE} produces parameter values that differ significantly from those computed by \texttt{ProSpect}. This suggests that differences arise not just from the choice of SPS model but also from the underlying methodologies and assumptions of each code which cannot be fully untangled.

The BPASS model yields 10 Myr timescale SFR medians similar to the BC03 model, though estimates a $\sim$40\% wider IQR. This does not significantly impact the median SFR$_{100}$ or stellar mass. We observe a slight increase in median \siion\, of 0.06 dex with a narrower IQR. While this is an expected result of potentially increasing the lifetimes of massive stars, the extent of this impact is below that found in similar studies. \cite{Emami2020} estimate a $\sim$0.15 dex difference for an EoR analog sample using a constant SFH. Using the BlueTides simulation, \cite{Wilkins2016a} found that the BPASS model resulted in a higher value of \siion\ than the BC03 model, by a factor of 0.14 dex and 0.36 dex for the single-star and binary models, respectively.

Comparing the \texttt{BEAGLE} BC03 and BC16 models, we observe only marginal differences in median SFR$_{10}$ (\SF=0.08 dex), SFR$_{100}$ (\SF=0.3 dex) and stellar mass $\Delta$\M= 0.1 dex. The median \siion\, are also found to be only 0.05 dex apart, though a slightly narrower range of values is found using the BC16 model. Speculating on the source of this requires a deeper analysis of each model assumption.

\subsection{Dependence of \siion\ on Dust attenuation model}\label{subsec:Dust}
Dust attenuation models, used alongside re-emission models in SED fitting, describe how UV and optical light is absorbed and re-emitted by dust. They account for extinction, the line-of-sight dimming of light caused by dust properties (e.g., chemical composition, grain size), column density, and the spatial distribution of stars and dust \citep{Salim2020}. “Correcting for dust attenuation of stellar light is essential for estimating intrinsic stellar properties and has been shown to play a more significant role at high redshifts than previously anticipated \citep{Akins2023, Bisigello2024}. Any method used to model \siion\, requires the dust-corrected (i.e., unattenuated) UV luminosity at 1500\AA. As a result, all such estimates are intrinsically dependent on the choice of dust attenuation model. 

The \cite{Calzetti2000} dust attenuation model and the SMC extinction curve  \citep{Zubko1999} represent two extremes of dust attenuation behavior. The SMC model, derived from the Small Magellanic Cloud, features a steep extinction slope, while the Calzetti model shows a shallower curve typical of more metal-rich systems. Steep extinction curves like the SMC’s are generally associated with lower-metallicity, high-redshift populations \citep{Reddy2015}. In contrast, the shallower Calzetti curve is linked to higher-metallicity, low-redshift populations, though the geometry of stars and dust is also a major influencing factor \citep{Calzetti2012}. The Charlot and Fall (CF00, \citealt{Charlot2000}) and \texttt{BEAGLE}'s Universal model \citep{Chevallard2013} are both two-component dust models. They separately treat attenuation from the diffuse ISM, typically resembling a Calzetti-like curve, and from stellar birth clouds, which follow an SMC-like curve. This approach accounts for the additional attenuation experienced by young stars still embedded in their birth environments. The resulting effective attenuation curve is shaped by the balance between the two components and typically lies between the Calzetti and SMC curves (see \citealt{Battisti2020}). The Universal model further differs by implicitly considering the inclination angle of the galaxy into the attenuation estimate.

\begin{figure*}
    \centering
\begin{minipage}{0.70\textwidth}
        \centering
        \includegraphics[width=\textwidth]{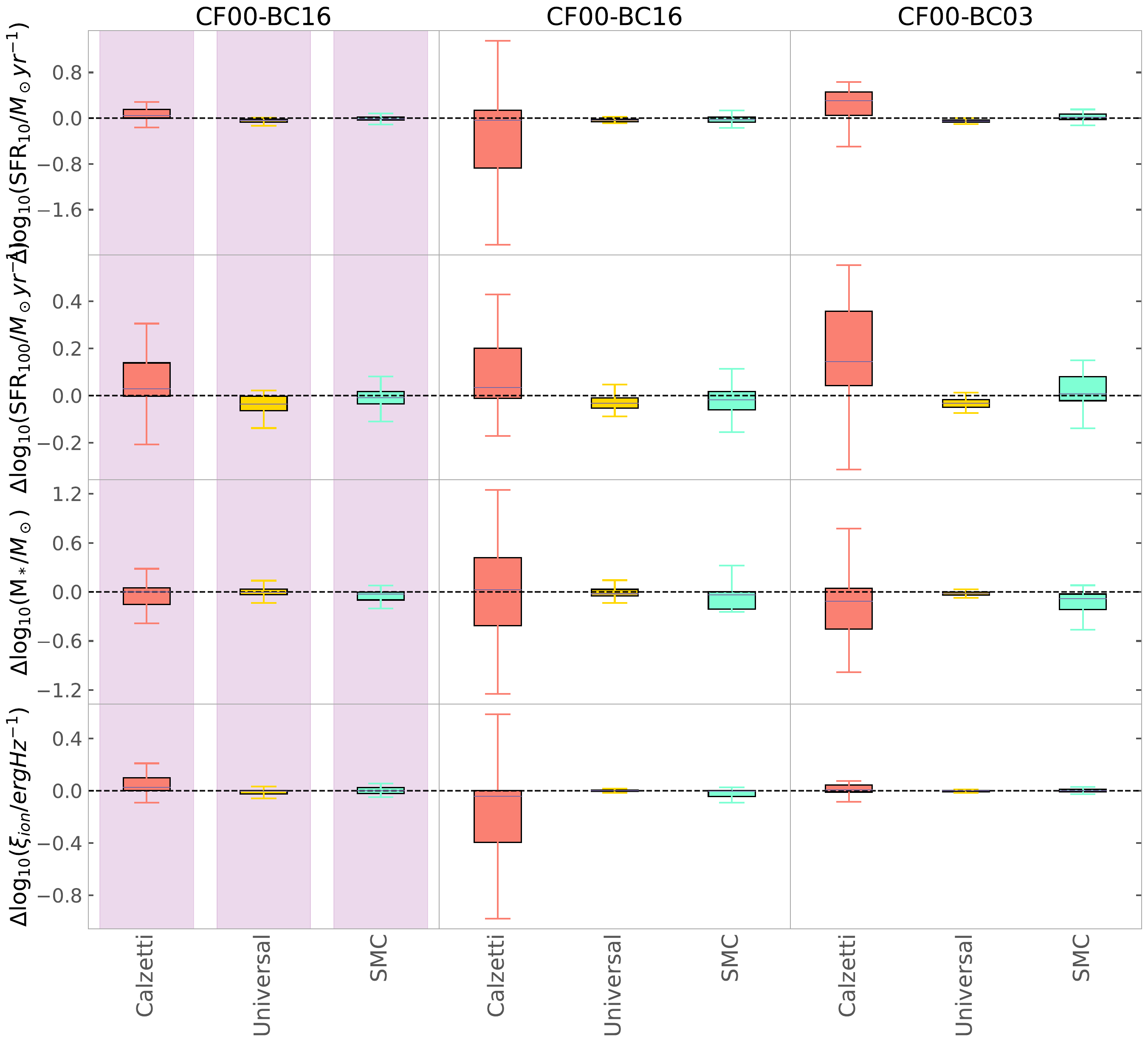}
    \end{minipage}

\begin{minipage}{0.4\textwidth}
        \centering
        \includegraphics[width=\textwidth]{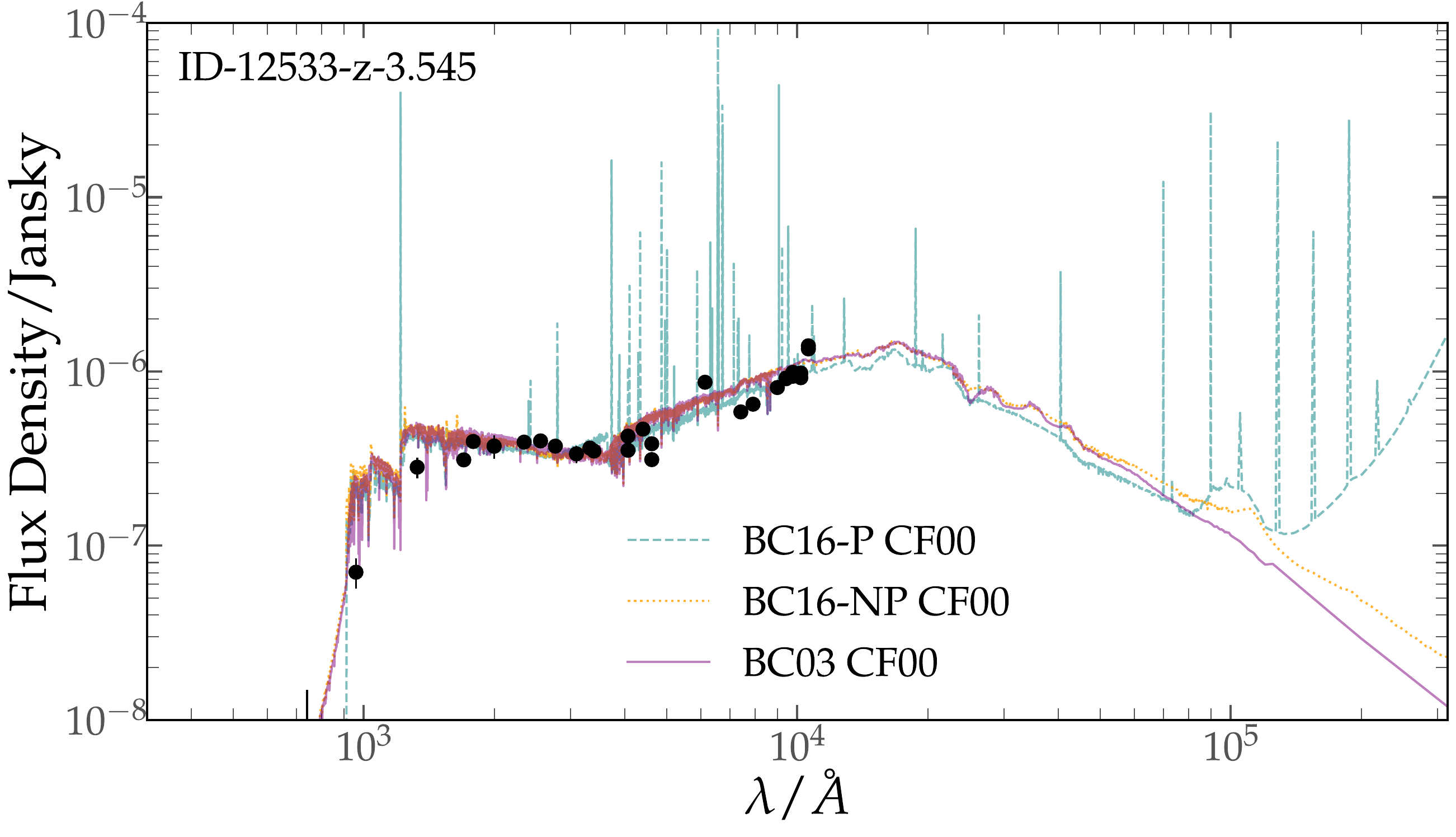}
    \end{minipage}\hfill
    \begin{minipage}{0.4\textwidth}
        \centering
        \includegraphics[width=\textwidth]{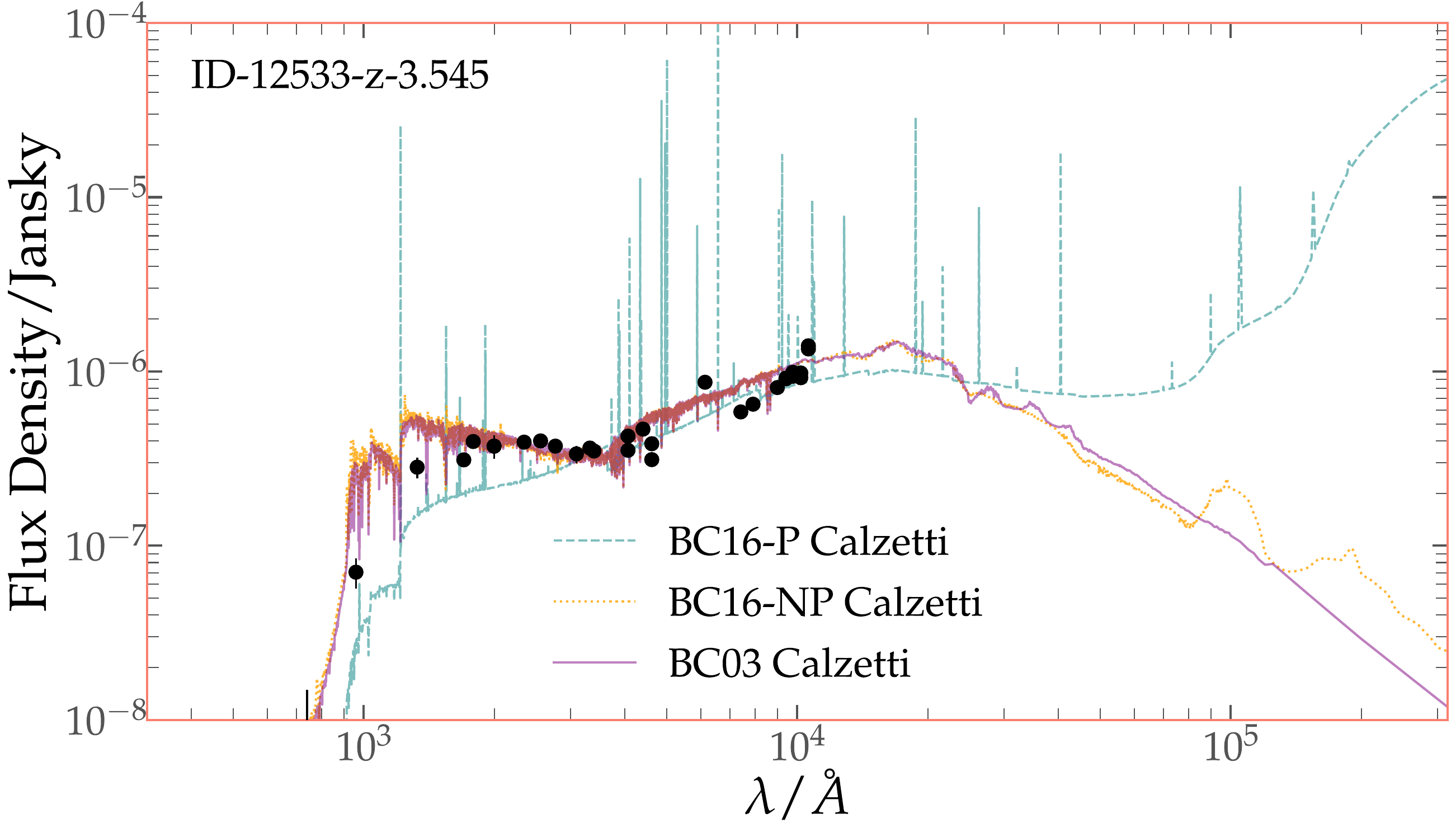}
    \end{minipage}

    \vspace{0.2em} 

    \begin{minipage}{0.4\textwidth}
        \centering
        \includegraphics[width=\textwidth]{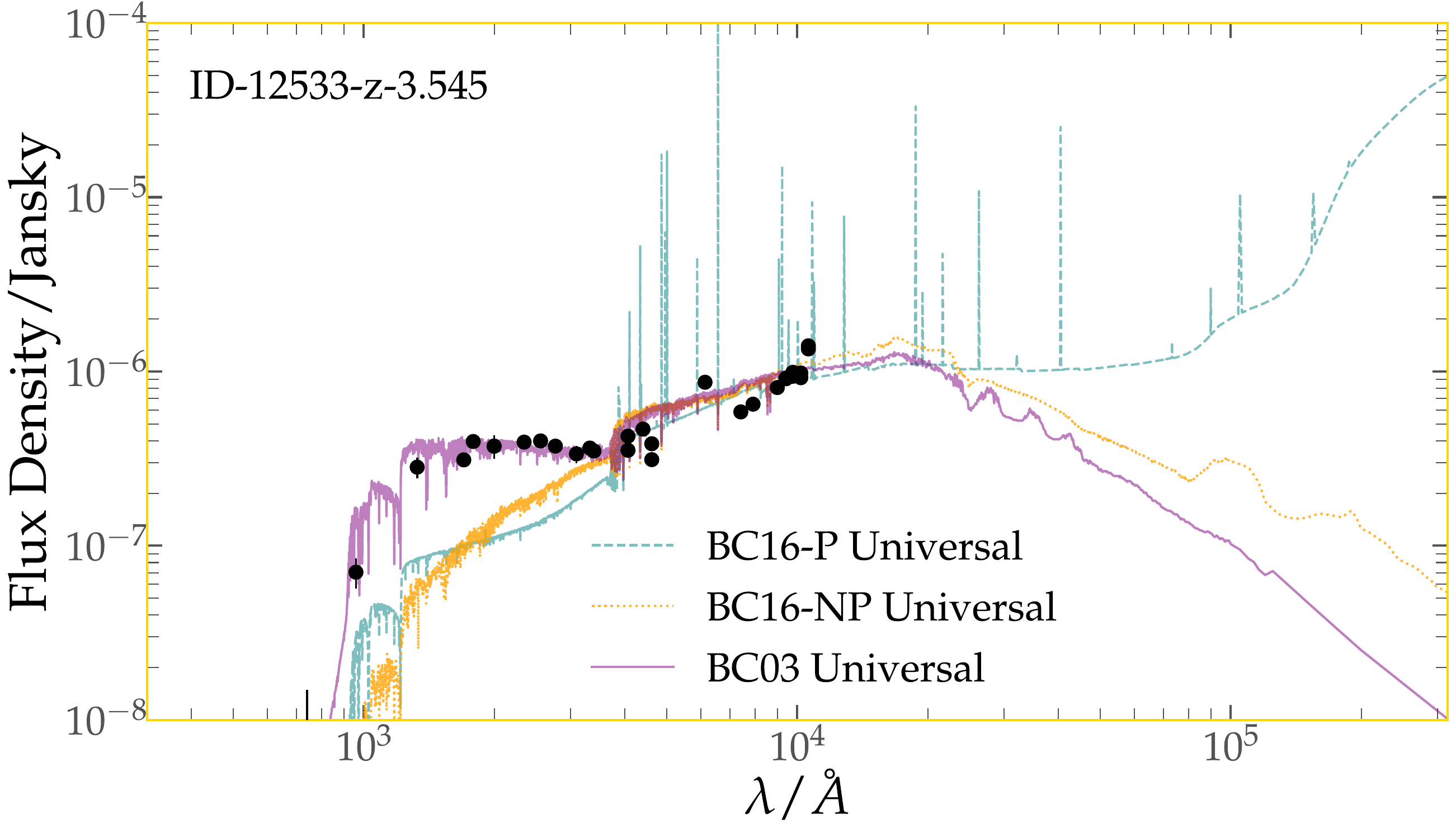}
    \end{minipage}\hfill
    \begin{minipage}{0.4\textwidth}
        \centering
        \includegraphics[width=\textwidth]{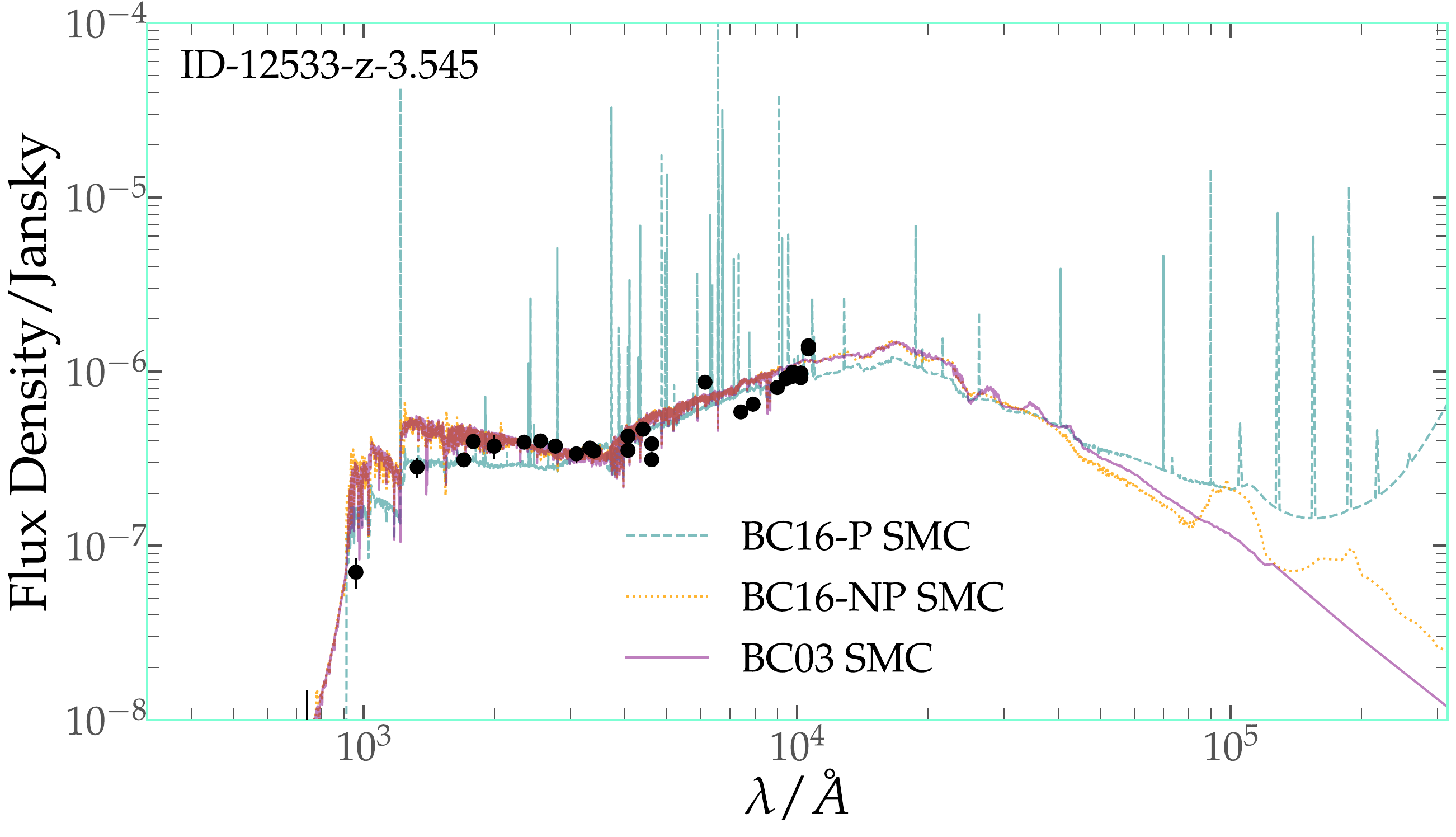}
    \end{minipage}

    \caption{Dependence of stellar parameters on dust model in \texttt{BEAGLE}; purple shading indicates models including photoionization (CLOUDY \cite{Ferland2017}) All parameters are shown relative to the Charlot and Fall 2000 (CF) dust model and using the same SPS model (indicated at top). Top panel. CF-BC16 (left). CF dust model +BC16 SPS including photoionization,  subtracted from the Calzetti, Universal and SMC dust models also including photoionization and BC16. CF-BC16 (middle). CF dust model +BC16 SPS but excluding photoionization, subtracted from the equivalent setup for the Calzetti, Universal and SMC dust models. CF-BC03 (right). CF dust model +BC03 SPS excluding photoionization subtracted from dust models using the same setup. Boxes show the IQR (25–75\%), with the median in grey; whiskers extend to 1.5$\times$IQR. Bottom grid. Bottom panels show the corresponding SEDs; P/NP denote models with/without photoionization (corresponding to the first and second columns of the Top Panel) and the BC03 result (third column of the top panel). We note that these galaxies aren't dusty \citep{Jaiswar2024}, and that these differences could be further exaggerated in other samples. We find that our galaxies only significantly deviate in stellar parameters using the Calzetti model without photoionization.}
    \label{fig:Dust}
\end{figure*}

To isolate the effects of dust attenuation models and minimize potential confounding from different SPS or photoionization models, this analysis  (Figure \ref{fig:Dust}) focuses solely on results from the  \texttt{BEAGLE} SED fitting code and is presented in three parts. Results are presented as differences in parameter values relative to those obtained with the CF00 model, which is the default in \texttt{BEAGLE}. That is, for each galaxy, the median and IQR values reflect the difference between the chosen dust model and the CF00 model (with a consistent SPS/photoionization setup). This choice is motivated by the added flexibility of a two-component model compared to the empirical SMC and Calzetti models, and the significance of birth cloud attenuation in EELGs. This section also considers the impact of photoionization modeling as it influences the underlying continuum fit.

As noted in section \ref{sec:data}, previous studies \citep{Jaiswar2024} have shown that this sample exhibits low dust attenuation. Therefore, these results may not fully encapsulate the potential influence of the chosen models on a general population. However, as EoR analogues, these results would be representative for comparative studies of bright sources in the early universe. 

Using the BC16 SPS model (Charlot \& Bruzal in prep)  with CLOUDY photoionization modeling (BC16-P; Figure \ref{fig:Dust}), we find that each of the Calzetti, Universal and SMC models yield nearly identical median values and inter-quartile ranges for the SFR$_{10}$, SFR$_{100}$ and M$_*$. Relative to CF00, the Calzetti model estimates slightly elevated 75th percentile SFR$_{10}$ (\SF=0.14 dex) and SFR$_{100}$ (\SF=0.6 dex ). It also predicts a lower 25th percentile stellar mass ($\Delta\log$(M/M$_\odot$)=0.15 dex below CF00) and a higher 75th percentile \siion\, (0.07 dex above CF00). However, these differences are small in contrast to those made without inclusion of photoionization modeling. 

With the exclusion of a photoionization model (BC16-NP), previously minor differences become more pronounced. Despite similar median SFR estimates, the Calzetti model yields a significantly lower 25th percentile  SFR$_{10}$ (\SF=0.34 dex) compared to CF00. In contrast, the  75th percentile SFR$_{100}$ is higher by \SF=0.14 dex. Taking these results together suggests the SFH has a diminished ``burst" in the Calzetti prescription. The Calzetti model estimates a stellar mass IQR that is shifted by up to 0.4 dex relative to CF00. Notably, the 25th percentile \siion\, 25th percentile is 0.36 dex lower. This tendency of lower \siion\, combined with the Calzetti model's frequent use at low redshift could bias redshift evolution measurements in comparative studies. The SMC model differs insignificantly from CF00, suggesting that attenuation is primarily driven by the birth cloud component rather than the diffuse ISM. . 

In practice, most galaxy evolution studies use the BC03 SPS model, making it important to assess the impact of dust attenuation models specifically within this framework. The previous section enables this by isolating the effects of the SPS model change, without introducing additional complications from photoionization modeling, as would typically occur when using BC16. Compared to CF00, the Calzetti model consistently estimates galaxies to be more massive by 0.55 dex and more star-forming, with median increases of 0.28 dex in SFR$_{10}$ and 0.13 dex in SFR$_{100}$. This implies a bluer underlying flux density. Despite this, the median \siion\, differs by only 0.02 dex between the models, suggesting that the additional mass could be ascribed to older stars rather than heavily obscured massive stars. These differences are not reflected in the other dust models, which produce more consistent results with each other. 

Given the redshift range and the low-dust, low-metallicity nature of this sample, the Calzetti attenuation model is likely suboptimal for these EELGs. However, the increasing evidence for early quenching complicates assumptions about which attenuation models are appropriate at a given epoch \citep{Forrest2020,delaVega2025}.

\begin{figure*}[!htbp]
    \centering
    \includegraphics[width=0.8\linewidth]{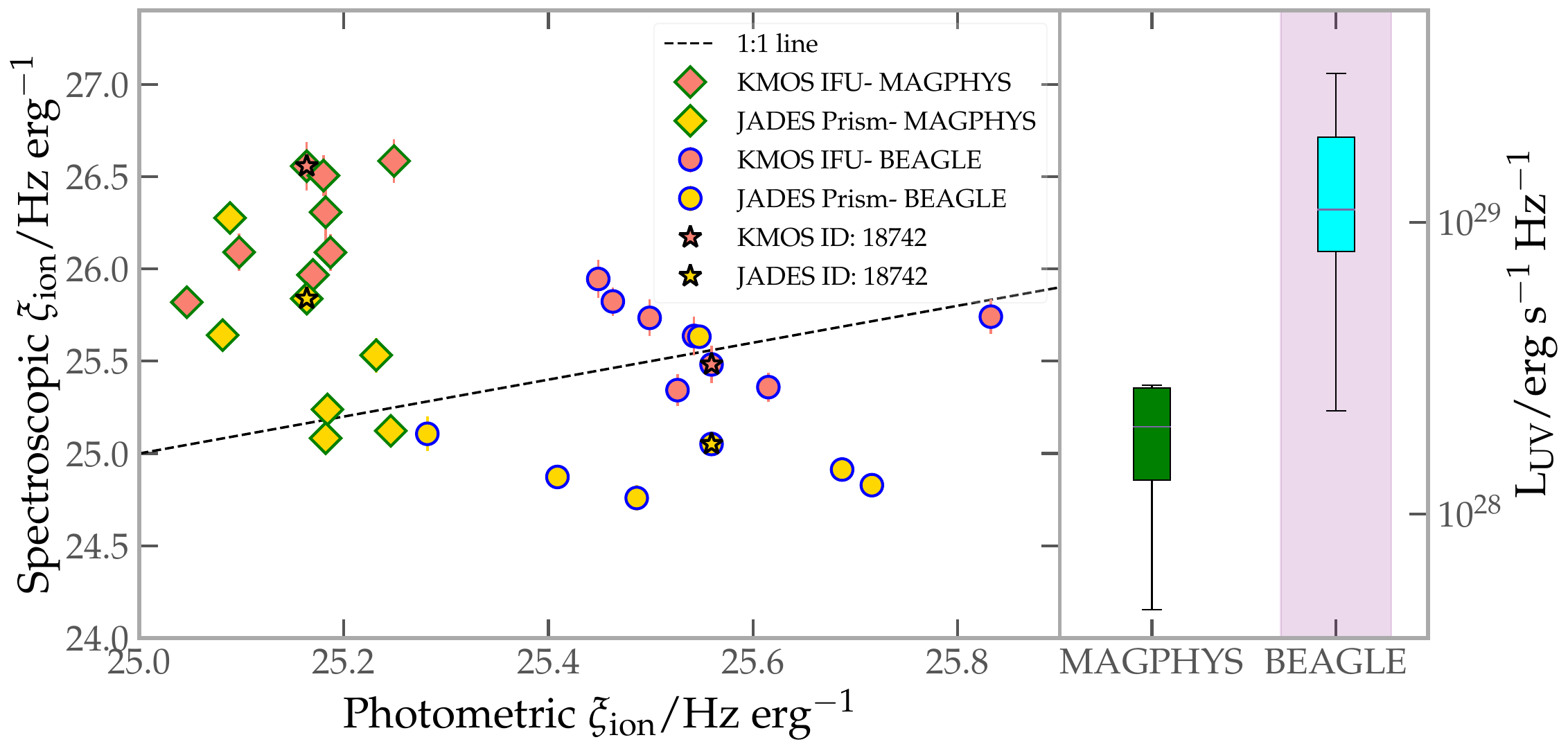}
    \includegraphics[width=0.8\linewidth]{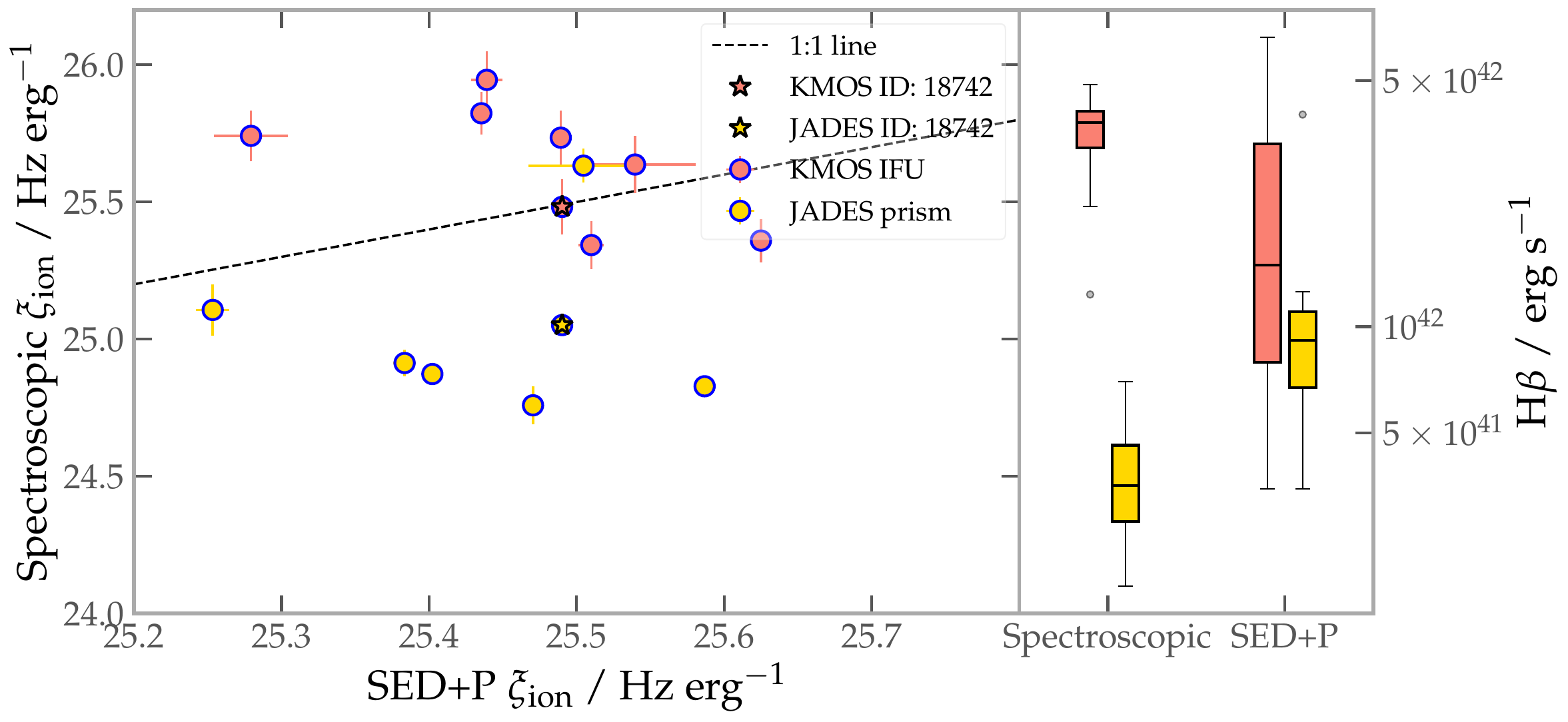}
    \includegraphics[width=0.8\linewidth]{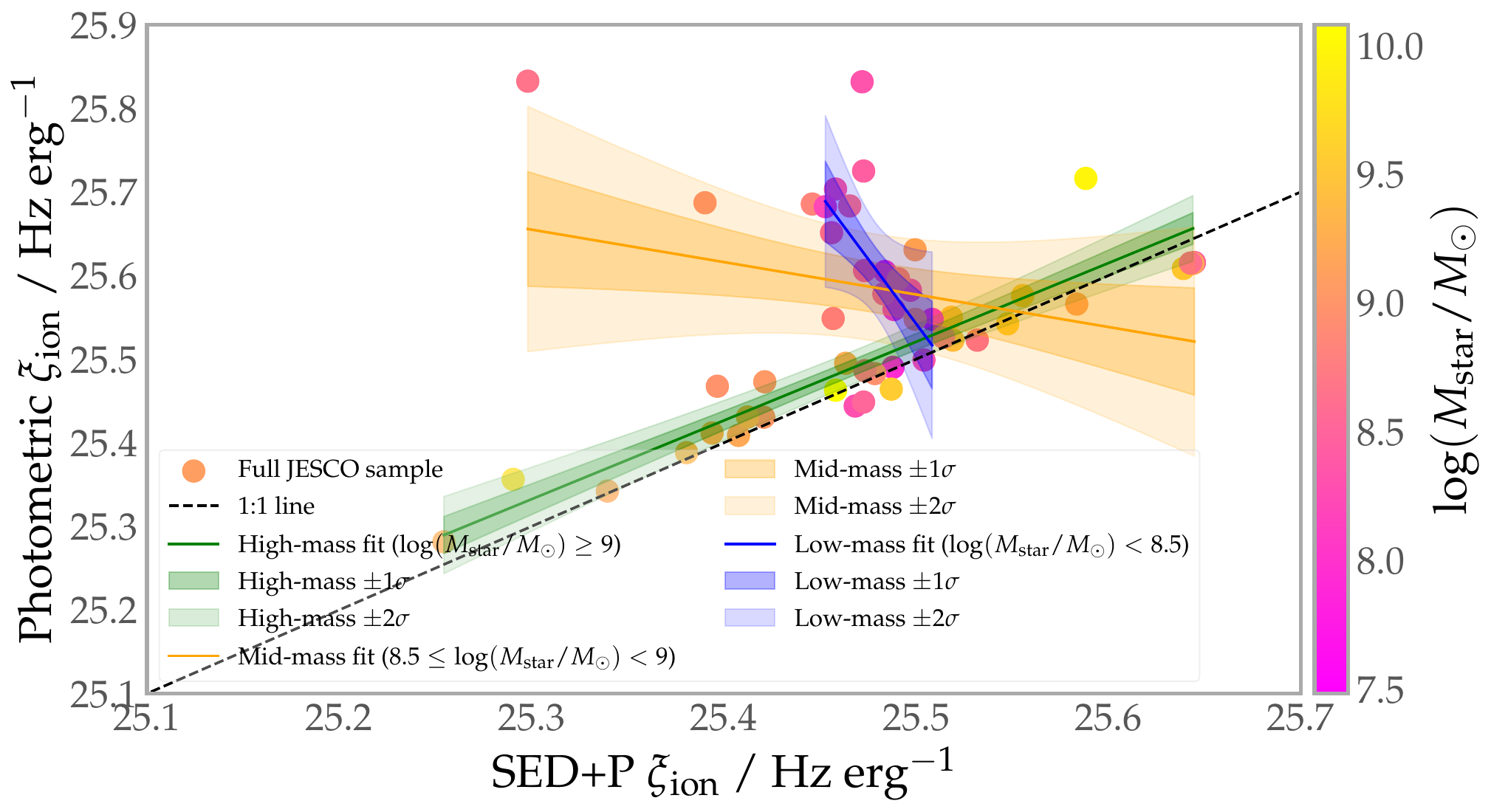}

    \caption{(Top) Spectroscopic \siion\, using VLT KMOS (red) and JWST JADES NIRSpec (gold) vs integrated photometry \siion .  $\mathrm{L_{UV}(1500)}$ IQRs indicated for \texttt{MAGPHYS} (diamonds) and \texttt{BEAGLE} (circles). (Middle) Spectroscopic \siion\, vs the photoionization model (SED+P) \siion\ with the H$\beta$ flux IQRs from the KMOS (red) and NIRSpec (gold) compared to the corresponding SED+P model estimate. (Bottom) Photoionization model \siion\, vs integrated photometry \siion\, with 1-2$\sigma$ ranges binned by M$_*$. A 1:1 black dotted line is given, and the same source (ID 18742) measured by each instrument is indicated by a star. Models are configured to their default settings (see Table \ref{tab:Default}). We find no correlation between methods relying on spectra vs those on photometry with slight improvements using photoionization modeling. We also find the same source measured by each instrument to vary in spectroscopic \siion\, by $\sim$0.5 dex, and see a notable mass dependence of the correlation between the integrated photometry \siion\, and the SED+P \siion .}
    \label{fig:specphot}
\end{figure*}

\subsection{Dependence of \siion\, on Technique}\label{subsec:spectrophotometry}
So far we have explored the model dependence of \siion\, when derived from photometry alone. This is crucial for understanding large-scale trends,as it leverages the broad availability of photometric data and allows for self-consistent modeling of dust correction and LyC production through SFH and SPS models. However, combining this method with spectroscopic measurements of nebular lines such as  H$\alpha$ or H$\beta$ offers a powerful cross-check: while photometry constrains stellar and dust parameters, spectroscopy directly probes ionized gas regions around massive stars. This spectroscopic derivation relies primarily on the assumptions of negligible LyC escape and minimal pre-nebular dust absorption—both of which are reasonable for the low-dust systems in our sample \citep{Smith2022}.

The integration of photoionization modeling into SED fitting offers additional insights . The aforementioned determination of short timescale ($<$10 Myr) star formation and accurate mass estimates for EELGs are among these, but the estimation of emission line flux also provides an opportunity to ``spectroscopically" determine \siion\, \citep{Simmonds2023} without the limiting statistical significance factor. This method avoids explicit assumptions about LyC escape but is sensitive to the parameterization and physics of the photoionization model used.

Nebular emission fluxes can be estimated from photometry by measuring the excess flux in a filter contaminated by emission lines—without requiring explicit photoionization modeling. This technique has been applied to derive \siion\,, in previous studies, such as  \cite{Bouwens2015}. However, photoionization-based SED fitting integrates emission lines into the broader modeling framework: it accounts for dust attenuation, isolates blended lines such as \nii\,and is self-consistent with the underlying stellar population and SFH assumptions. 

The extraction of H$\alpha$ luminosity for this purpose has been explored \citep{Simmonds2023}, though this method has not been explicitly validated in the way explored here. To assess the accuracy of different \siion\, derivation techniques, we compare the three approaches identified in section \ref{subsec:siionmethod}. Using spectra for a subsample of EELGs observed with the KMOS and NIRSpec instruments (described in section \ref{subsec:spectroscopy}), we compare the SED+P (Spectral Energy Distribution with Photoionization) technique to a true spectroscopically derived \siion\, and consider also the comparison to the photometric integration derivation. All experiments use either \texttt{BEAGLE} or \texttt{MAGPHYS} in their “default” configurations and adopt their respective $\mathrm{L_{UV}(1500)}$ normalization factors (see Table \ref{tab:Default}). Experiments using the SED+P method rely entirely on \texttt{BEAGLE} as it requires the CLOUDY photoionization model.       

Comparing the photometric integration to spectroscopic values (Figure \ref{fig:specphot} top panel) reveals that the \texttt{BEAGLE} model yields the closest agreement with spectroscopically derived values, whereas \texttt{MAGPHYS} systematically underestimates most of the tested sample. As described in sections \ref{subsec:Default} and \ref{subsec:Bursts}, \texttt{BEAGLE} incorporates photoionization-informed SFRs by adjusting the SFH in the final 10 Myr, often resulting in a strong recent burst for EELGs. The resulting $\mathrm{L_{UV}(1500)}$ value is 1.5-7$\times$ greater than that determined by \texttt{MAGPHYS} and a disproportionately larger integrated LyC region. The discrepancies in $\mathrm{L_{UV}(1500)}$ and the integrated LyC between the burst implementations of each code are primary drivers for the differences in \siion\, where the integrated LyC region fluctuates by more than an order of magnitude (10-15$\times$), which more than counteracts the increased $\mathrm{L_{UV}(1500)}$. {Physically, the larger recent SFH (i.e. the burst), increases the population of surviving massive stars, which can then contribute to both the LyC and $\mathrm{L_{UV}(1500)}$. We find that the more model dependent LyC increases by a larger margin than the photometrically constrained $\mathrm{L_{UV}(1500)}$ even when comparing only \texttt{BEAGLE} with and without a burst.}  

It is important to note that the spectroscopic H$\beta$ values have not been corrected for dust here, as the Balmer decrement is inaccessible for this sample using the KMOS instrument, which potentially exacerbates the discrepancy between the relative model agreements to the spectroscopic benchmark. Additionally, the limited number of sources with available spectroscopic data constrains our ability to generalize these findings, and while \texttt{BEAGLE} appears to reproduce spectral values more accurately in this sample, broader conclusions require a larger and more uniformly observed sample.   

A notable point is also the serendipitous presence of galaxy ID 18742 in both the JADES prism spectra and the KMOS IFU, which reveals the instrumental dependence of \siion\, as well. Prism spectra are noted to be lower than grism values by an average of 10\%  \citep{Bunker2023}, and the KMOS user manual specifies a 5-20\% uncertainty in the absolute flux calibration. The IFU measurements minimize flux losses by design, which points to how instrumental systematics can propagate into \siion\, discrepancies. 

An equivalent comparison of the photoionization (SED+P) and spectroscopic methods (Figure\ref{fig:specphot} middle) reveals a significant scatter and little correlation between the two. The photoionization model tends to predict H$\beta$ fluxes that lie intermediate between those measured by the prism and IFU, but the overlap is weak. This mismatch is not unexpected, given the  the modeled flux discrepancies and the uncorrected instrumental uncertainties (which are not reflected in the current error bars). As such, the apparent disagreement may in part be reconciled by improved treatment of systematic errors. Nevertheless, the small sample size with matched photometric and spectroscopic observations remains a key limitation, restricting the robustness of this validation exercise. 

An informative comparison is made between the photometric integration and SED+P methods (Figure \ref{fig:specphot} bottom). Logically, each of these methods derives the LyC from the same model, and where one integrates this directly into the \siion\, estimate, the other first redistributes the flux into various emission lines to model the nebular interaction. We find that the degree to which the two methods agree depends strongly on galaxy mass. More massive galaxies tend to lie close to the 1:1 relation, whereas lower-mass galaxies systematically diverge from it.{ This bimodality is consistent with the burst measure $\mathrm{\frac{SFR_{10}}{SFR_{100}}}$ (see Appendix Figure \ref{fig:appendix_a}) and the mass weighted age. Galaxies with high SED+P \siion\, generally exhibit stronger recent bursts, and the offset from the 1:1 relation becomes most pronounced when the stellar mass is concentrated in very young populations. In more massive systems, the LyC output appears to be reprocessed into Balmer-line emission more consistently, largely independent of burst strength. By contrast, low-mass galaxies with very young stellar populations show a  relative deficit in H$\alpha$ compared to their LyC output, implying that the SED+P derived \siion\ is particularly sensitive to how the model treats the youngest stellar populations and their nebular emission.}

{This behavior is likely driven by the imposed constraint SFR$_{10} \leq 10 \times$ SFR$_{100}$, which disproportionately limits the formation of massive stars in low-mass galaxies. As a result, these systems cannot sustain sufficiently strong bursts and instead encounter a ceiling in this parameter space. Large bursts in these galaxies therefore shift the stellar population to ages up to two orders of magnitude younger than in more massive bursting systems, amplifying the offset from the 1:1 relation. The resulting bimodality is thus most likely attributable to the internal treatment of young stellar populations within \texttt{BEAGLE}, rather than to a physical origin.} 

{We investigated whether this feature could instead arise from the photoionization grid by examining the ionization parameter and gas-phase metallicity, neither of which shows a clear correlation with the bimodality. We also tested models with nonzero LyC escape fraction, rather than the fiducial assumption of $f_{\rm esc}=0$. Allowing LyC escape reduces the bimodality, but at the cost of introducing a larger systematic offset between the two methods (Appendix Figure \ref{fig:appendix_b}). This behavior is consistent with fewer ionizing photons being available to power nebular emission, while the fits compensate by favoring younger and more strongly burst-dominated populations. More broadly, this highlights the risk of interpreting differences between model-dependent \siion\ estimators as direct constraints on LyC production or photoionization physics without carefully accounting for the assumptions built into the fitting framework. The priors adopted here are motivated by recent high-redshift \siion\ studies of low-mass galaxies using the same model \citep{Boyett2024}, and may therefore have broader implications for how ionizing parameters are inferred in similar analyses.}

{We also note a small offset between the median of the high-mass galaxies and the 1:1 line, with the photometric-integration method yielding slightly larger LyC fluxes on average. This may reflect the same mass-dependent trend seen across the sample, although the limited upper end of our mass range prevents a stronger test using a higher mass cut.}

\begin{figure*}[!htbp]
    \centering
    \includegraphics[width=0.98\linewidth]{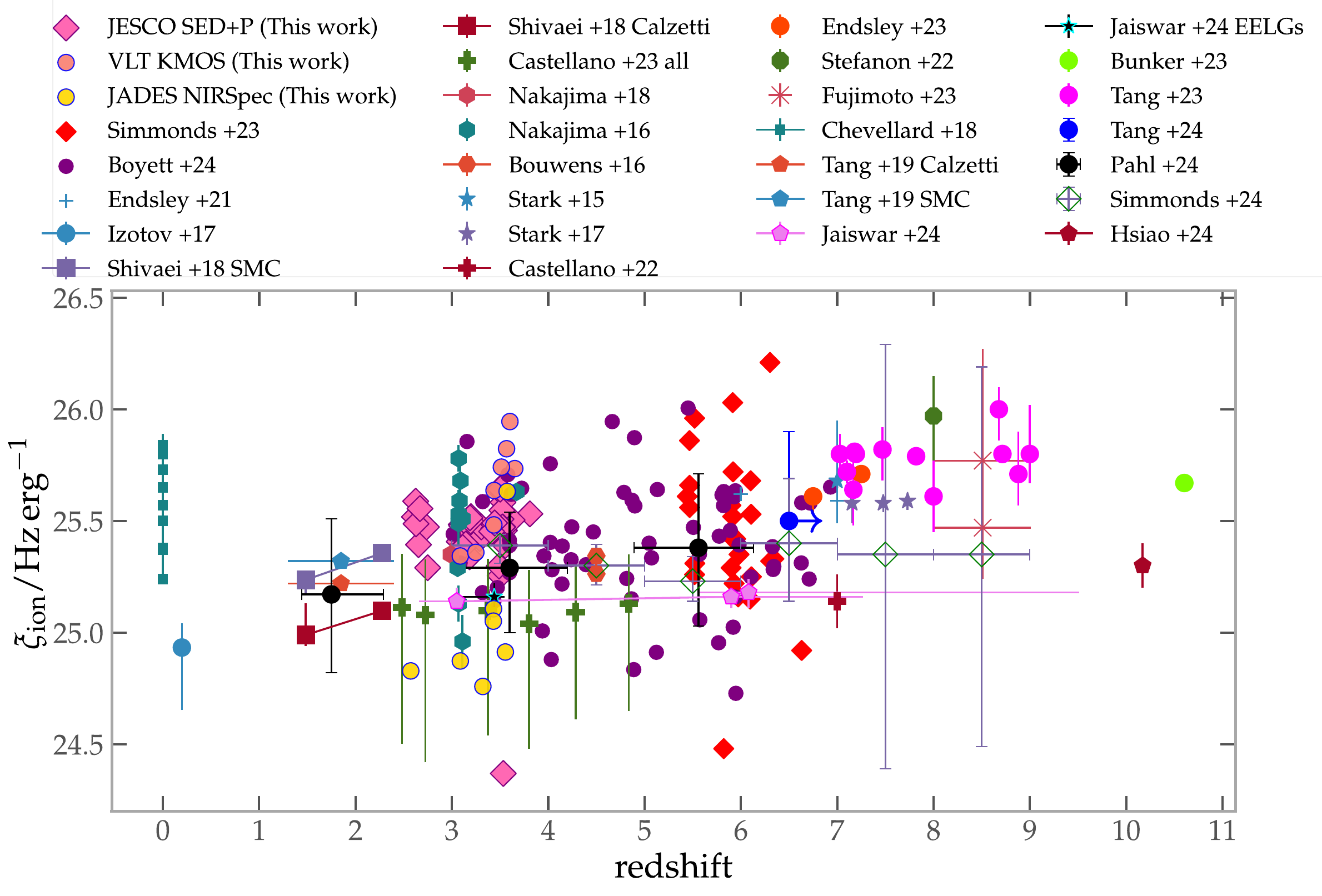}
    \includegraphics[width=0.48\linewidth]{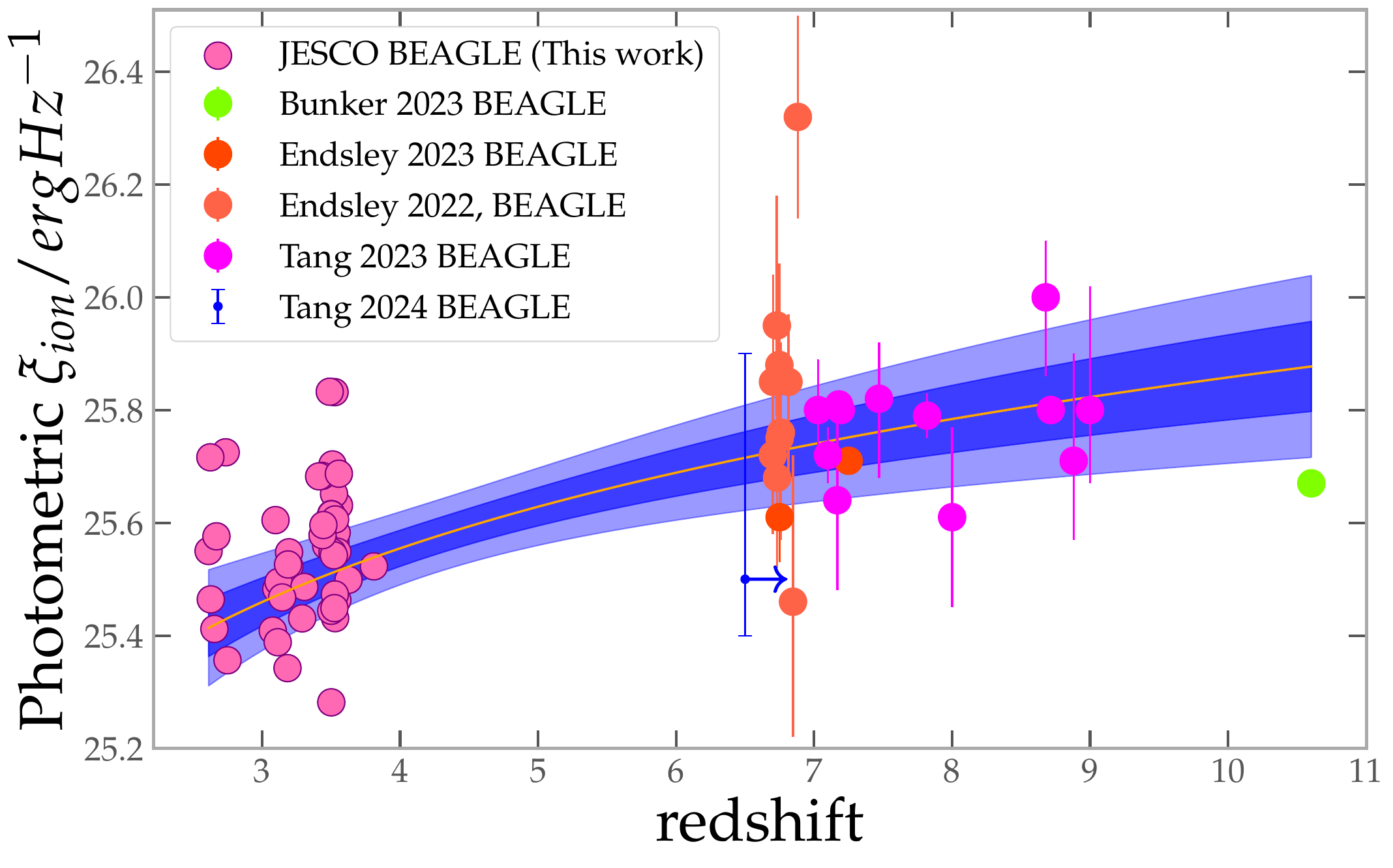}
    \includegraphics[width=0.48\linewidth]{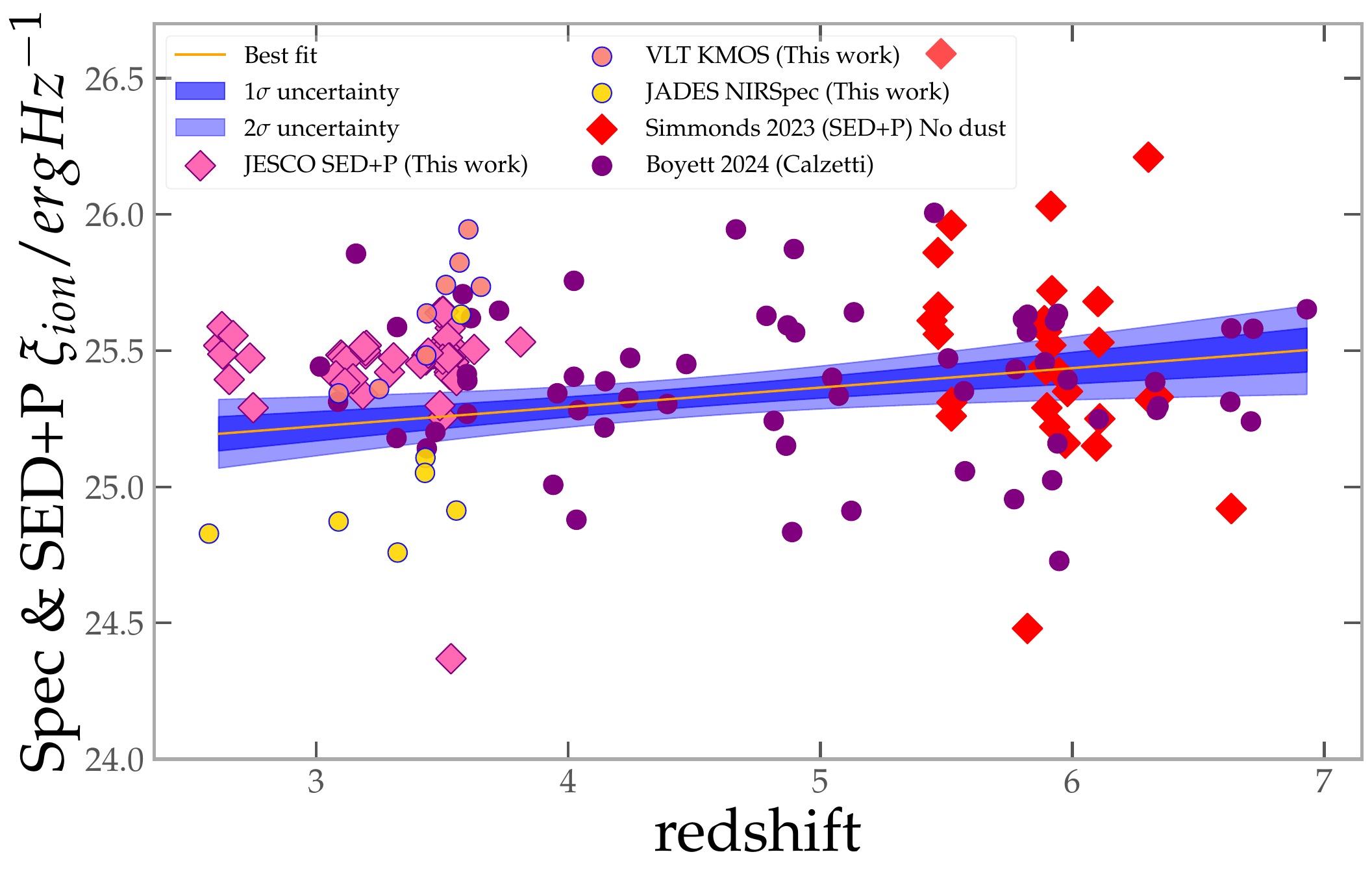}
    
    \caption{Redshift evolution of the \siion\, parameter. (Top) Indiscriminant selection of studies using both spectroscopic, photometric and the SED+P method for sources varying in sample selection choices and size \citep{Bouwens2015,Stark2015,Nakajima2016,Stark2017,Izotov2017,Chevallard2018,Nakajima2018,Shivaei2018,Tang2019,Castellano2022,Stefanon2022,Fujimoto2023,Castellano2023,Simmonds2024,Jaiswar2024,Hsiao2024,Pahl2024}. (Bottom left) Selection of only studies using the photometric integration method and the \texttt{BEAGLE} code \citep{Endsley2021,Endsley2023,Bunker2023b,Tang2023,Tang2024}. (Bottom right) Selection of only spectroscopic and SED+P method studies\citep{Simmonds2023,Boyett2024} and only those which derive the \siion\ and $\mathrm{L_{UV}(1500)}$ using \texttt{BEAGLE}. Blue shaded regions indicate the 1-2$\sigma$ errors around the median shown as an orange line. Discrepancies in selected dust models or methods are noted. We find a clear evolution in only the \texttt{BEAGLE} photometric integration \siion\, and no correlation with redshift in the \texttt{BEAGLE} SED+P and Spectroscopic methods (bottom right), or the indiscriminant methods sample (top). Determining the significance of the evolution requires additional data coverage using one method.}
    \label{fig:Redshift}
\end{figure*}

\section{Evolution of \siion\ with Redshift}\label{sec:Redshift Evolution}

Understanding the correlation of the \siion\, parameter to physical characteristics can help us better grasp this otherwise unintuitive parameter in more tangible terms, and several studies have explored such correlations \citep{Castellano2023,Jaiswar2024,Boyett2024}. While evidence suggests that \siion\, correlates with sSFR and anti-correlates with stellar mass \citep{Castellano2023,Jaiswar2024}, the redshift evolution remains unclear. For instance, \cite{Matthee2016} defines an increasing \siion\ with increasing redshift using a selection of H$\alpha$ and Ly$\alpha$ emitters between $2.2<z<5$, whereas other studies fail to reproduce any trend \citep{Castellano2023,Simmonds2024}. 

Such discrepancies may arise from biases related to sample selection and the over-representation of bright or extreme sources. A recognized way to mitigate this is to bin galaxies by stellar mass, which helps control for mass-dependent effects. We will argue that it is equally important to compare results from studies that adopt similar models and methodologies, as significant offsets can be introduced by variations in SPS, SFH, dust attenuation models, and data types.

To illustrate this point, Figure \ref{fig:Redshift} (Top) uses an indiscriminant sample of galaxies from redshift 0-11, consisting of EELGs and more normal/quiescent systems. This compilation uses each of the techniques described in section \ref{subsec:spectrophotometry} and employs various dust treatments, SPS models or SFHs. Unsurprisingly, no discernible trend with redshift emerges (no significant Spearman correlation {p$>$ 0.05}). This ambiguity could reflect true intrinsic scatter, but it may also result from mass incompleteness, methodological inconsistency, or model dependence in the \siion\, determinations. 

In contrast, upon isolating only samples sharing consistent methodologies (Figure \ref{fig:Redshift}  Bottom left) and using the same model, we derive a clearer trend (Spearman correlation of 0.56, $p<0.9e^{-7}$). The sample selection here is also more similar, with our JESCO sample aiming to mimic the EELG properties of EoR galaxies. The best fit logarithmic relation is $0.08\times ln(z)+25.10$. Provided the model dependence of this \siion\, determination, a possible caveat is that this evolutionary trend is a product of the \texttt{BEAGLE} model itself and testing this would require a large redshift coverage of similar samples using another model. However, these wider studies do not exist for the \texttt{ProSpect} or \texttt{MAGPHYS} models currently. Furthermore, culling the literature sample to this subset limits the redshift coverage and sample size, which could skew the observed trend. 

A similar restriction applied to spectroscopic or SED+P samples (Figure \ref{fig:Redshift} bottom right) reveals a large scatter (no significant Spearman correlation {p$>$0.05}) and a much more modest increase of \siion\, with redshift. This sample also relies on the \texttt{BEAGLE} model for the $\mathrm{L_{UV}(1500)}$ normalization factor, which may again impose model-related limitations. The reduced dynamic range in redshift and the small sample size further weaken the significance of any potential trend.

The contrasting conclusions of the redshift evolution depictions in Figure \ref{fig:Redshift} reflect the difference made by methodology, sample selection and selected models to the \siion\,evolution problem. It is entirely plausible that the depicted evolution using the photometric method is an artifact of the chosen models in each sample. Simultaneously, it is also plausible that spectroscopic selections targeting bright or extreme galaxies could limit the significance of a discernible trend. The assumptions made by each technique, such as the \fesc=0 in the spectroscopic derivation or the modeled O star population in the photometric method must also be critically evaluated. 

In summary, while each method suffers its own caveats, meaningful comparisons of \siion\,, across different studies require consistent modeling frameworks and well-matched sample selections. Even the conclusions drawn within a single study must be properly tempered with its own limitations. Without standardization, conflicting results are likely to persist. 

\section{Conclusion}\label{sec:conclusions}
We summarise our main conclusions below, noting that all numerical values apply specifically to extreme emission line galaxies (EELGs) at z$\sim$3, and should not be generalized to broader galaxy populations: 
\begin{enumerate}
     \item The default configurations of different SED fitting codes can significantly alter the conclusions drawn on the inferred stellar parameters and particularly the \siion. Considerations must be made when reporting these parameters and comparing them to those determined by other studies. 
    \item The chosen star formation history model is a key determining factor for calculating stellar parameters. Including flexibility by having more complex functions with more independent parameters in the SFH is reflected in the broader distribution of each measured parameter.  
    \item The implementation of star formation bursts was found to be the most influential individual component affecting the magnitude of \siion\,. The inclusion of the CSFH free parameter in a bursty population increases the calculated \siion\ by $\sim$0.3 dex relative to its removal. The additional massive stars contribute more to the LyC budget (10-15$\times$) than to the $\mathrm{L_{UV}(1500)}$ (1-7$\times$), increasing \siion.  
    \item While the choice of SPS model has limited impact on most stellar parameters, it does modestly affect \siion\,. We find that models increasing the lifespan or prevalence of massive stars have slightly elevated \siion\,.  
    \item The chosen dust model appears to only marginally impact the determination of stellar parameters for our (non-dusty) sample, with the largest influence being in absence of a concurrent photoionization model. This mostly applies to the choice of the Calzetti model vs SMC-like extinctions with up to a 0.4 dex discrepancy in \siion\, based on this choice. With an uncertain population of high redshift dusty sources, this choice becomes less clear.
    \item The SED+P method described here provides a theoretical alternative to spectroscopic measurements of \siion\,{ however, based on this analysis the approach suffers from significant SFR model biases}. With considering also the substantial systematic uncertainties in the spectra themselves, validation of this approach would benefit from application to a larger sample and with a single spectroscopic instrument.
    \item The photometric integration and SED+P methods have an {unexpected} mass-dependent relation in their agreement to one-another. {We found that this is an artifact of the model SFR creating extremely young stellar populations that dominated the ionizing thoroughput without converting efficiently to H$\alpha$ emission.} Their agreement supports the photometric integration method's validity for massive galaxies, given that the SED+P method reasonably reconstructs nebular emission lines.
    \item Including studies that do not use similar methodologies or models washes out any underlying evolution in correlations to redshift {under their differing assumptions and model biases. Using a single method across a wide redshift coverage in both directions is necessary for stricter conclusions, however, the biases in the chosen method must be considered by comparative studies such as this. }
\end{enumerate}

\section{Acknowledgements}

This research was partly supported by the Australian Research Council Centre of Excellence for All Sky Astrophysics in 3 Dimensions (ASTRO 3D), through project number CE170100013. CMT was supported by an ARC Future Fellowship under grant FT180100321.
\\The International Centre for Radio Astronomy Research (ICRAR) is a Joint Venture of Curtin University and The University of Western Australia, funded by the Western Australian State government.
RJ would like to thank Andy Bunker for helpful discussions and Jessica E Thorne for their help with ProSpect. 

\FloatBarrier
\appendix
\label{sec:appendix}

\begin{figure}[b]
    \centering
    \includegraphics[width=0.75\linewidth]{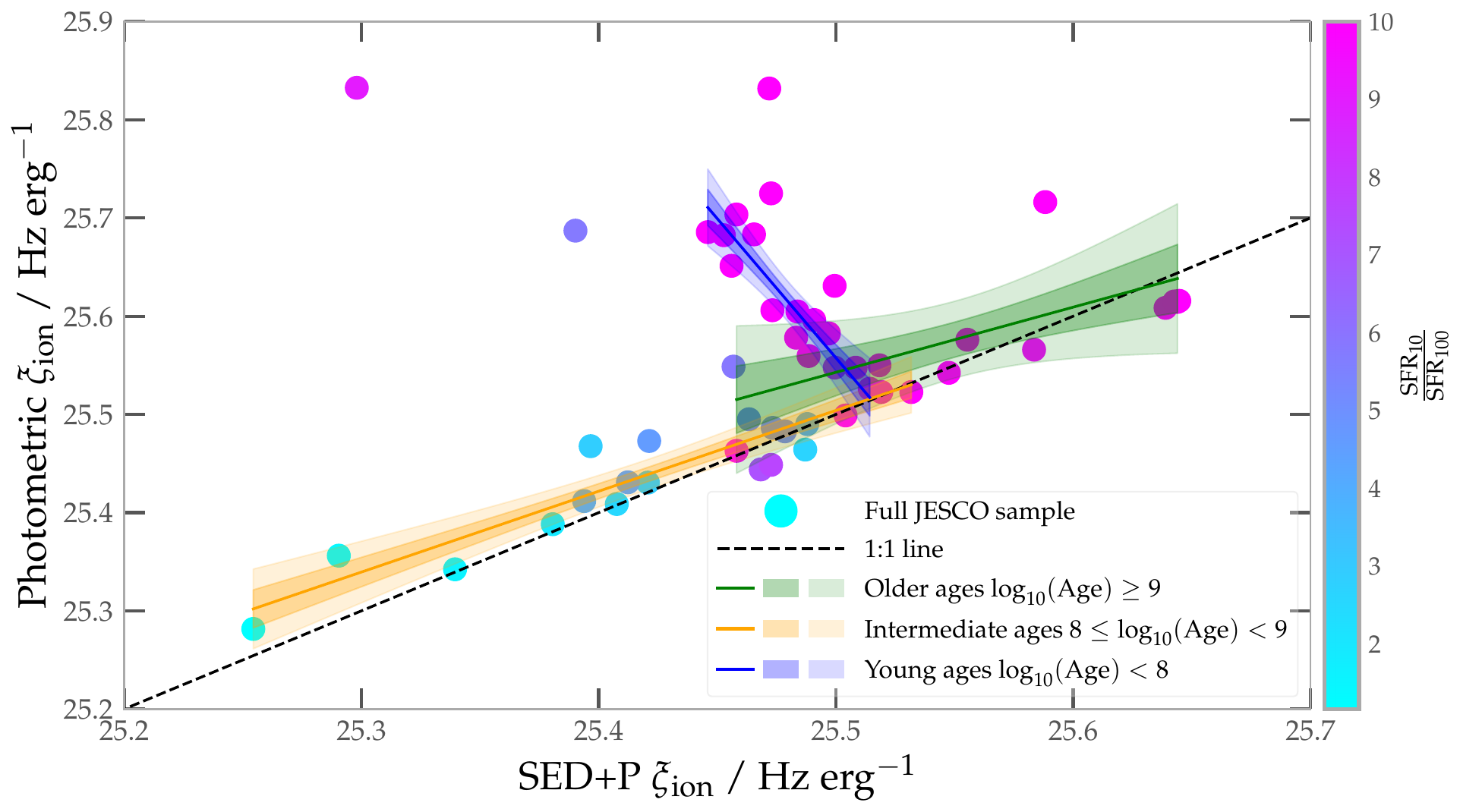}
    \caption{\texttt{BEAGLE} photoionization model \siion\, (SED+P) vs integrated photometry \siion\, following Figure \ref{fig:specphot}, colored by burst measure $\mathrm{\frac{SFR_{10}}{SFR_{100}}}$. Medians and 1-2$\sigma$ shading for intervals in bins of mass-weighted age. Galaxies diverging from the 1:1 relation are primarily low-mass, burst-dominated systems with young mass-weighted ages, whereas older galaxies remain closer to the 1:1 relation even at similarly high burst measures. This behavior is consistent with a limitation in the SFR and photoionization modeling, in which the ionizing radiation produced by the youngest stellar populations is not converted into nebular emission with the same efficiency across the sample. In these systems, the inferred \siion\, from the SED+P method is offset because a smaller fraction of the LyC output effectively powers recombination-line emission, causing the low-mass, young population to separate from the main relation.}
    \label{fig:appendix_a}
\end{figure}

\begin{figure}
    \centering
    \includegraphics[width=0.60\textwidth]{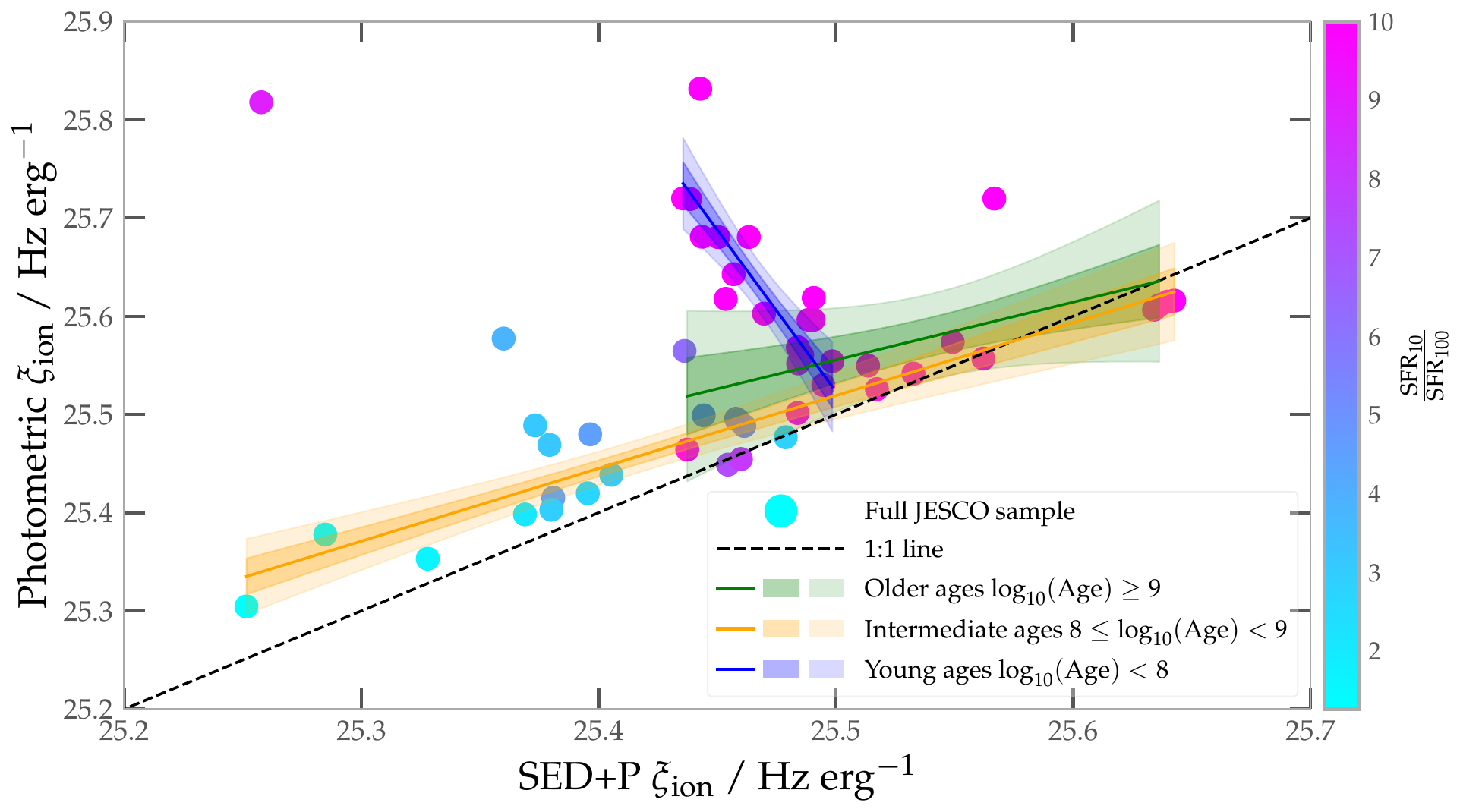}\\
    \includegraphics[width=0.60\textwidth]{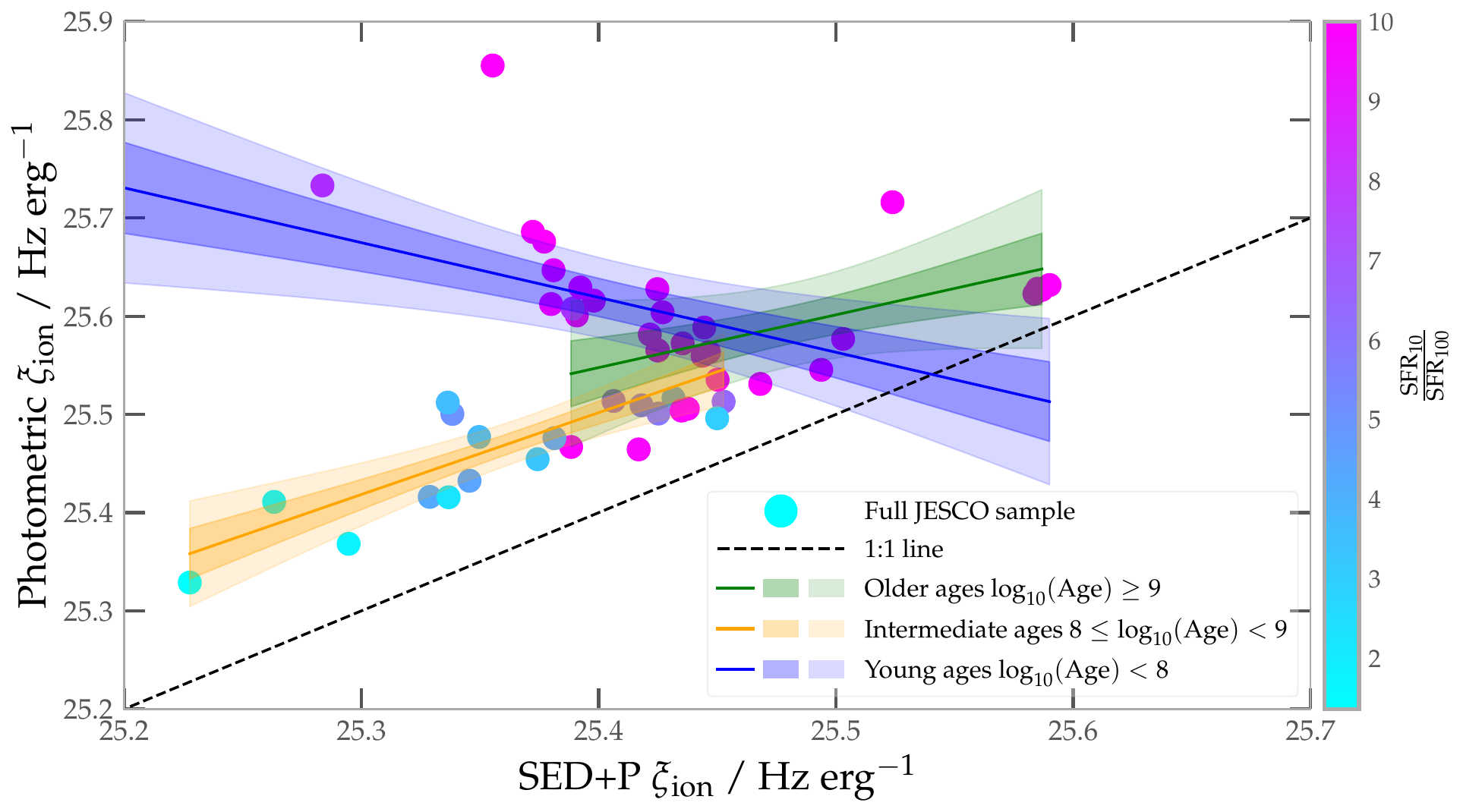}\\
    \includegraphics[width=0.60\textwidth]{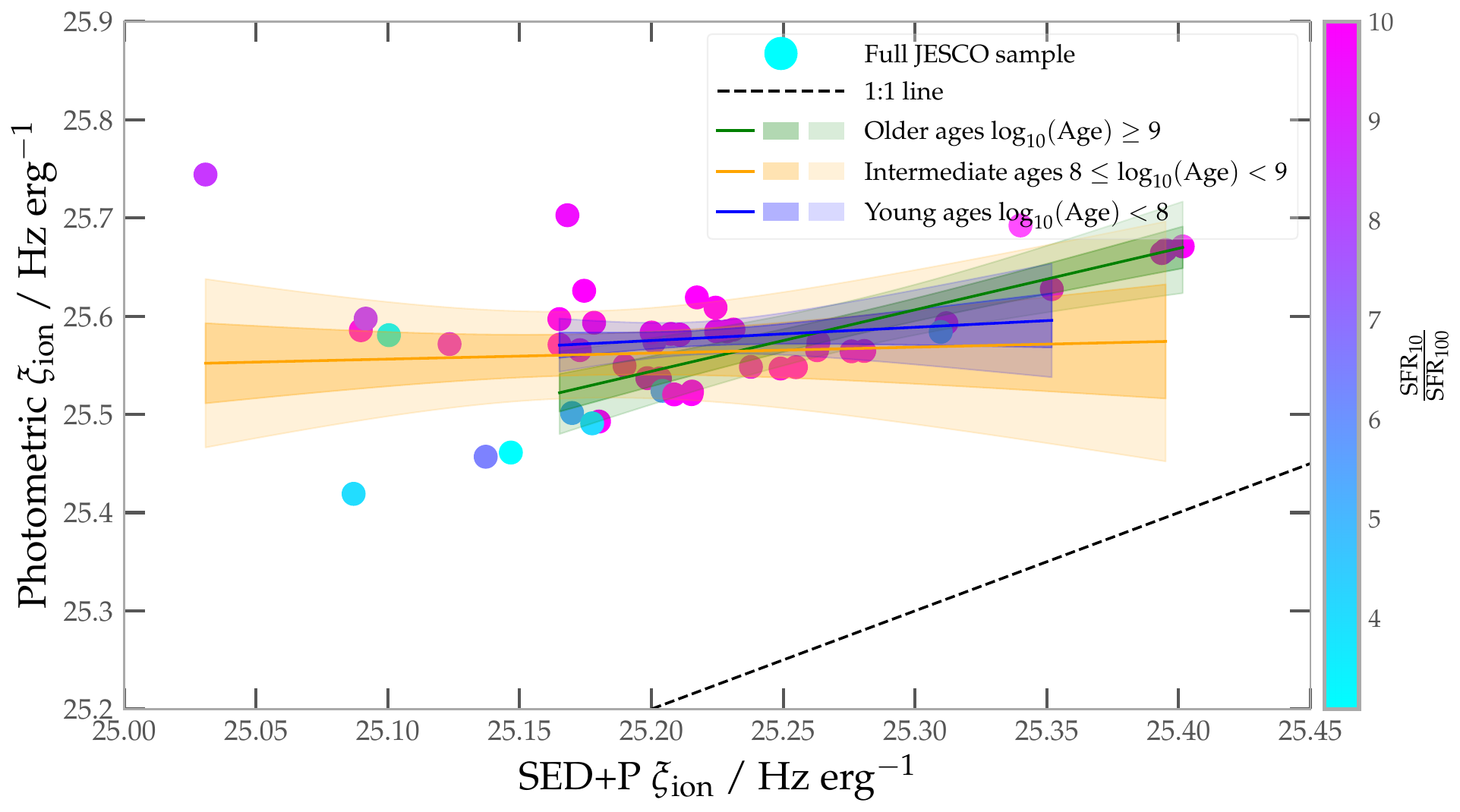}
    \caption{\texttt{BEAGLE} photoionization model \siion\, (SED+P) vs integrated photometry \siion\, following Figure \ref{fig:specphot}, colored by burst measure $\mathrm{\frac{SFR_{10}}{SFR_{100}}}$. Medians and 1-2$\sigma$ shading for intervals in bins of mass-weighted age. The top, middle, and bottom panels assume LyC escape fractions of (fitted) 1–5\%, (fixed) 15\%, and (fixed) 50\%, respectively.The bimodality described in section \ref{subsec:spectrophotometry} weakens with increasing LyC escape fractions , while the relation as a whole deviates systematically from the 1:1 line. This trend is consistent with a larger fraction of LyC photons escaping before being converted into nebular emission. Larger escape fractions reduce the need for a distinct population of galaxies with an apparently anomalous nebular conversion efficiency. In particular, the low-mass, young systems that previously occupied a separate high \siion\, mode instead converge toward the main population as the models uniformly favor younger ages and more extreme recent bursts. The increased escape fraction forces the models to adopt a larger $\mathrm{L_{UV}(1500)}$ relative to the LyC budget, such that the observed photometry is reproduced with younger, more burst-dominated stellar populations while the inferred ionizing photon production efficiency decreases.}
    \label{fig:appendix_b}
\end{figure}

The bimodality of the high and low mass subsets in Figure \ref{fig:specphot} is explored in terms of the mass-weighted ages and burst measures in Figure \ref{fig:appendix_a}. We entertain a varied ionizing photon escape fraction as a potential reconciliation to the Figure \ref{fig:specphot} bimodality in Figure \ref{fig:appendix_b}. We provide the format of the derived photometry in Table \ref{tab:JESCO1}.

\clearpage

\startlongtable
\begin{deluxetable}{lll}
\tablecaption{JESCO Derived Photometry\label{tab:JESCO1}}
\tablehead{
 \colhead{Units} & \colhead{Label} & \colhead{Description}
}
\startdata
---  &    \phm{e\_}zfourge                  & ZFOURGE identifier\tablenotemark{1} \\
deg  &    \phm{e\_}RAdeg                    & Right Ascension, decimal degrees, ZFOURGE (J2000)\tablenotemark{1} \\
deg  &    \phm{e\_}DEdeg                    & Declination, decimal degrees, ZFOURGE (J2000)\tablenotemark{1} \\
Jy   & \phm{e\_}flux-wfc3-uvis-f336w     & Flux density, HST/WFC3 UVIS F336W \\
Jy   &  e\_flux-wfc3-uvis-f336w     & Uncertainty in flux-wfc3-uvis-f336W \\
Jy   & \phm{e\_}flux-acs-wfc-f435w       & Flux density, HST/ACS WFC F435W \\
Jy   &  e\_flux-acs-wfc-f435w       & Uncertainty in flux-acs-wfc-f435W \\
Jy   & \phm{e\_}flux-acs-wfc-f475w       & Flux density, HST/ACS WFC F475W \\
Jy   &  e\_flux-acs-wfc-f475w       & Uncertainty in flux-acs-wfc-f475W \\
Jy   & \phm{e\_}flux-wfc3-uvis-f606w     & Flux density, HST/WFC3 UVIS F606W \\
Jy   &  e\_flux-wfc3-uvis-f606w     & Uncertainty in flux-wfc3-uvis-f606W \\
Jy   & \phm{e\_}flux-acs-wfc-f606w       & Flux density, HST/ACS WFC F606W \\
Jy   &  e\_flux-acs-wfc-f606w       & Uncertainty in flux-acs-wfc-f606W \\
Jy   & \phm{e\_}flux-acs-wfc-f775w       & Flux density, HST/ACS WFC F775W \\
Jy   &  e\_flux-acs-wfc-f775w       & Uncertainty in flux-acs-wfc-f775W \\
Jy   & \phm{e\_}flux-wfc3-uvis-f814w     & Flux density, HST/WFC3 UVIS F814W \\
Jy   &  e\_flux-wfc3-uvis-f814w     & Uncertainty in flux-wfc3-uvis-f814W \\
Jy   & \phm{e\_}flux-acs-wfc-f814w       & Flux density, HST/ACS WFC F814W \\
Jy   &  e\_flux-acs-wfc-f814w       & Uncertainty in flux-acs-wfc-f814W \\
Jy   & \phm{e\_}flux-acs-wfc-f850lp      & Flux density, HST/ACS WFC F850LP \\
Jy   &  e\_flux-acs-wfc-f850lp      & Uncertainty in flux-acs-wfc-f850LP \\
Jy   & \phm{e\_}flux-nircam-j-f090w      & Flux density, JWST/NIRCam JADES F090W \\
Jy   &  e\_flux-nircam-j-f090w      & Uncertainty in flux-nircam-j-f090W \\
Jy   & \phm{e\_}flux-wfc3-uvis-f850lp   & Flux density, HST/WFC3 UVIS F850LP \\
Jy   &  e\_flux-wfc3-uvis-f850lp   & Uncertainty in flux-wfc3-uvis-f850LP \\
Jy   & \phm{e\_}flux-wfc3-ir-f105w       & Flux density, HST/WFC3 IR F105W \\
Jy   &  e\_flux-wfc3-ir-f105w       & Uncertainty in flux-wfc3-ir-f105W \\
Jy   & \phm{e\_}flux-nircam-j-f115w      & Flux density, JWST/NIRCam JADES F115W \\
Jy   &  e\_flux-nircam-j-f115w      & Uncertainty in flux-nircam-j-f115W \\
Jy   & \phm{e\_}flux-wfc-ir-f110w        & Flux density, HST/ACS IR F110W \\
Jy   &  e\_flux-wfc-ir-f110w        & Uncertainty in flux-wfc-ir-f110W \\
Jy   & \phm{e\_}flux-wfc3-ir-f125w       & Flux density, HST/WFC3 IR F125W \\
Jy   &  e\_flux-wfc3-ir-f125w       & Uncertainty in flux-wfc3-ir-f125W \\
Jy   & \phm{e\_}flux-wfc3-ir-f140w       & Flux density, HST/WFC3 IR F140W \\
Jy   &  e\_flux-wfc3-ir-f140w       & Uncertainty in flux-wfc3-ir-f140W \\
Jy   & \phm{e\_}flux-nircam-j-f150w      & Flux density, JWST/NIRCam JADES F150W \\
Jy   &  e\_flux-nircam-j-f150w      & Uncertainty in flux-nircam-j-f150W \\
Jy   & \phm{e\_}flux-wfc3-ir-f160w       & Flux density, HST/WFC3 IR F160W \\
Jy   &  e\_flux-wfc3-ir-f160w       & Uncertainty in flux-wfc3-ir-f160W \\
Jy   & \phm{e\_}flux-nircam-j-f182m      & Flux density, JWST/NIRCam JADES F182M \\
Jy   &  e\_flux-nircam-j-f182m      & Uncertainty in flux-nircam-j-f182M \\
Jy   & \phm{e\_}flux-nircam-f182m        & Flux density, JWST/NIRCam FRESCO F182M \\
Jy   &  e\_flux-nircam-f182m        & Uncertainty in flux-nircam-f182M \\
Jy   & \phm{e\_}flux-nircam-j-f200w      & Flux density, JWST/NIRCam JADES F200W \\
Jy   &  e\_flux-nircam-j-f200w      & Uncertainty in flux-nircam-j-f200W \\
Jy   & \phm{e\_}flux-nircam-f210m        & Flux density, JWST/NIRCam FRESCO F210M \\
Jy   &  e\_flux-nircam-f210m        & Uncertainty in flux-nircam-f210M \\
Jy   & \phm{e\_}flux-nircam-j-f210m      & Flux density, JWST/NIRCam JADES F210M \\
Jy   &  e\_flux-nircam-j-f210m      & Uncertainty in flux-nircam-j-f210M \\
Jy   & \phm{e\_}flux-nircam-j-f277w      & Flux density, JWST/NIRCam JADES F277W \\
Jy   &  e\_flux-nircam-j-f277w      & Uncertainty in flux-nircam-j-f277W \\
Jy   & \phm{e\_}flux-nircam-j-f335m      & Flux density, JWST/NIRCam JADES F335M \\
Jy   &  e\_flux-nircam-j-f335m      & Uncertainty in flux-nircam-j-f335M \\
Jy   & \phm{e\_}flux-nircam-j-f356w      & Flux density, JWST/NIRCam JADES F356W \\
Jy   &  e\_flux-nircam-j-f356w      & Uncertainty in flux-nircam-j-f356W \\
Jy   & \phm{e\_}flux-nircam-j-f410m      & Flux density, JWST/NIRCam JADES F410M \\
Jy   &  e\_flux-nircam-j-f410m      & Uncertainty in flux-nircam-j-f410M \\
Jy   & \phm{e\_}flux-nircam-j-f430m      & Flux density, JWST/NIRCam JADES F430M \\
Jy   &  e\_flux-nircam-j-f430m      & Uncertainty in flux-nircam-j-f430M \\
Jy   & \phm{e\_}flux-nircam-f430m        & Flux density, JWST/NIRCam FRESCO F430M \\
Jy   &  e\_flux-nircam-f430m        & Uncertainty in flux-nircam-f430M \\
Jy   & \phm{e\_}flux-nircam-f444w        & Flux density, JWST/NIRCam FRESCO F444W \\
Jy   &  e\_flux-nircam-f444w        & Uncertainty in flux-nircam-f444W \\
Jy   & \phm{e\_}flux-nircam-j-f444w      & Flux density, JWST/NIRCam JADES F444W \\
Jy   &  e\_flux-nircam-j-f444w      & Uncertainty in flux-nircam-j-f444W \\
Jy   & \phm{e\_}flux-nircam-f460m        & Flux density, JWST/NIRCam FRESCO F460M \\
Jy   &  e\_flux-nircam-f460m        & Uncertainty in flux-nircam-f460M \\
Jy   & \phm{e\_}flux-nircam-j-f460m      & Flux density, JWST/NIRCam JADES F460M \\
Jy   &  e\_flux-nircam-j-f460m      & Uncertainty in flux-nircam-j-f460M \\
Jy   & \phm{e\_}flux-nircam-f480m        & Flux density, JWST/NIRCam FRESCO F480M \\
Jy   &  e\_flux-nircam-f480m        & Uncertainty in flux-nircam-f480M \\
Jy   & \phm{e\_}flux-nircam-j-f480m      & Flux density, JWST/NIRCam JADES F480M \\
Jy   &  e\_flux-nircam-j-f480m      & Uncertainty in flux-nircam-j-f480M \\
---  &    zpk                    & ZFOURGE EASY photometric redshift\tablenotemark{1}  \\
\enddata
\tablenotetext{1}{Identifiers, positions, and photometric redshifts taken exactly from ZFOURGE \citep{Straatman2016}, corresponding to the 1997 release on their website.}
\tablecomments{Table \ref{tab:JESCO1} is published in its entirety in the electronic 
edition of the {\it Astrophysical Journal}.  A portion is shown here 
for guidance regarding its form and content.}
\end{deluxetable}

\FloatBarrier
\bibliography{example}{}
\bibliographystyle{aasjournal}

\end{document}